\documentclass[usenatbib]{mnras}
\usepackage[T1]{fontenc}
\usepackage{ae,aecompl}
\usepackage{siunitx}
\usepackage[fleqn]{amsmath}
\usepackage{amssymb}
\usepackage{wasysym}
\usepackage{booktabs}
\usepackage{tabularx}
\usepackage{graphicx}
\graphicspath{{figure/}}
\usepackage{rotating}
\usepackage{url}
\usepackage{color}
\newcommand{\Te}{T$_\text{e}$}
\newcommand{\trel}{$t_{2}$-$t_{3}$ relation\ }
\newcommand{\oiii}{[O{\sc iii}]\ }
\newcommand{\oii}{[O{\sc ii}]\ }
\newcommand{\nii}{[N{\sc ii}]\ }
\newcommand{\sii}{[S{\sc ii}]\ }

\newcommand{\Hii}{H{\sc ii}\ }

\title{New fully empirical calibrations of strong-line metallicity indicators in star forming galaxies}

\author[M. Curti et al.]{
M. Curti,$^{1,2}$\thanks{E-mail:mcurti@arcetri.astro.it}
G. Cresci,$^{2}$
F. Mannucci,$^{2}$
A. Marconi,$^{1}$
R. Maiolino$^{3,4}$
and S. Esposito$^{2}$
\\
$^{1}$Dipartimento di Fisica e Astronomia, Universit\`a di Firenze, Via G. Sansone 1, I-50019, Sesto Fiorentino (Firenze), Italy\\
$^{2}$INAF - Osservatorio Astrofisico di Arcetri, Largo E. Fermi 5, I-50125, Firenze, Italy\\
$^{3}$Cavendish Laboratory, University of Cambridge, 19 J. J. Thomson Ave., Cambridge CB3 0HE, UK\\
$^{4}$Kavli Institute for Cosmology, University of Cambridge, Madingley Road, Cambridge CB3 0HA, UK\\}

\begin{document}
\label{firstpage}
\pagerange{\pageref{firstpage}--\pageref{lastpage}}
\maketitle

\begin{abstract}
We derive new empirical calibrations for strong-line diagnostics of gas phase metallicity in local star forming galaxies by uniformly applying the Te method over the full metallicity range probed by the Sloan Digital Sky Survey (SDSS). To measure electron temperatures at high metallicity, where the auroral lines needed are not detected in single galaxies, we stacked spectra of more than 110000 galaxies from the SDSS in bins of log[O{\sc ii}]/H$\beta$ and log[O{\sc iii}]/H$\beta$. This stacking scheme does not assume any dependence of metallicity on mass or star formation rate, but only that galaxies with the same line ratios have the same oxygen abundance. We provide calibrations which span more than 1 dex in metallicity and are entirely defined on a consistent absolute Te metallicity scale for galaxies. We apply our calibrations to the SDSS sample and find that they provide consistent metallicity estimates to within 0.05 dex.
\end{abstract}
%
\begin{keywords}
galaxies: abundances -- galaxies: evolution -- ISM: abundances
\end{keywords}

\section{Introduction}

The accurate determination of gas phase metallicity represents a challenging topic for studies that aim at understanding the chemical evolution of galaxies. 
The metal content of a galaxy is regulated by complex interactions between physical processes occurring on different spatial and time scales: heavy elements produced by stellar activity contribute to the enrichment of the interstellar medium (ISM), while cosmological infall of pristine gas from the intergalactic medium (IGM) and outflows due to Active Galactic Nuclei (AGNs) and supernovae could dilute the ISM and at the same time trigger new star formation episodes \citep{Dave:2011aa}.
These processes directly impact the global baryon cycle and thus affect other physical quantities strictly related to the history of chemical enrichment in galaxies like stellar mass (M$_{\star}$) and star formation rate (SFR); therefore, relationships between these parameters and gas-phase metallicity are expected.
Indeed in the last decades strong observational evidences of a correlation between M$_{\star}$ and gas-phase metallicity (the so called mass-metallicity relation, M-Z) have been reported by several studies, 
both in the local Universe (e.g. \citealt{Tremonti:2004aa,Lee:2006aa,Liang:2007aa}) and at higher redshift, where signatures of a cosmic evolution have been found (e.g \citealt{Erb:2006aa, Maiolino:2008cv, Mannucci:2009aa, Zahid:2012aa, Cresci:2012aa, Troncoso:2014aa}).
Furthermore, \cite{Mannucci:2010gy} showed that the intrinsic scatter in the M-Z could be reduced when SFR is taken into account, introducing the concept of a Fundamental Metallicity Relation (FMR) that reduces the M-Z to a two-dimensional projection of a three dimensional surface. 
The FMR appear to be \textit{more fundamental} in the sense that it does not seem to present clear signs of evolution up to z $\sim 2.5$.
Even though the physical origin of these relations is still debated,
the knowledge of the main properties of the M-Z and the exact form of its dependence upon the SFR is important to investigate the processes regulating star formation and to assess the role of outflows in ejecting metals out of the galaxy \citep{Dave:2011aa,Lilly:2013aa,Dayal:2013aa}; this could provide crucial observational constraints for models aimed at reproducing the chemical evolution of galaxies across cosmic time.

Investigating the properties of these relationships and their redshift evolution requires precise and robust metallicity estimates. Since the scatter in the FMR is of the order of $0.05$ dex \citep{Mannucci:2010gy}, such a level of precision in metallicity determination would be desirable. 
There are several ways to measure abundances in galaxies, but unfortunately none of them is considered reliable or applicable over the whole metallicity range covered by large galaxy samples.
The most commonly used method relies on the determination of the electron temperature of the nebulae responsible for emission lines in galaxies: in fact, electron temperature is known to be strongly correlated with metallicity, such that higher metallicities are associated to lower \Te, because forbidden emission lines from metals are the primary coolants in \Hii regions. Electron temperatures can be inferred by exploiting the temperature sensitive auroral to nebular line ratios of particular ions (e.g. \oiii$\lambda 4363$/$5007$ is one of the most widely used); in fact, the atomic structure of these ions is such that auroral and nebular lines originate from excited states that are well spaced in energy and thus their relative level populations depend heavily on electron temperature.
This so called \Te\ method is widely accepted as the preferred one to estimate abundances since it is a direct probe of the processes that regulate the physics of ionized nebulae.
Unfortunately, auroral lines are weak in most of individual galaxy spectra, especially for metal rich objects, which typically prevents the \Te\ from being used method to determine abundances of metal enriched galaxies. 
A different technique is based instead on exploiting the ratio between oxygen and hydrogen recombination lines (RLs): since these lines show a very weak dependence on electron temperature and density (\citealt{Esteban:2009gu, Esteban:2014aa, Peimbert:2014aa}) this is probably the most reliable method beacuse is unaffected by the typical biases of the \Te\ method associated with temperature fluctuations. 
Typical discrepancies between \Te\ and RLs based abundances are found to be of the order 
of $0.2$-$0.3$ dex, with the first ones underestimating the latter (\citealt{Garcia-Rojas:2007aa, Esteban:2009gu}).   
Recently, \cite{Bresolin:2016aa} showed that metal RLs yield nebular abundances in excellent agreement with stellar abundances for high metallicity systems, while in more metal-poor environments they tend to underestimate the stellar metallicities by a significant amount. 
However, RLs are extremely faint (even hundred times fainter than oxygen auroral lines) and cannot be detected in galaxies more distant than a few kpc \citep{Peimbert:2007aa}.
For this reason, different methods have been developed to measure abundances in faint, distant 
and high metallicity galaxies.
In particular, it is known that some line ratios between strong collisionally excited lines (CELs) and Balmer lines show a dependence on metallicity, which can be either directly motivated or indirectly related to other physical quantities (e.g. the ionization parameter). 
Thus, it has been proposed that these line ratios could be calibrated against the oxygen abundance and used as metallicity indicators for galaxies in which the application of the \Te\ method is not possible due to the extreme faintness of auroral lines \citep{Pagel:1979pd,Alloin:1979lq}: these are referred to as the strong-line-methods. 
Calibrations can be obtained either empirically, for samples in which metallicity have been previously derived with the \Te\ method (e.g. \citealt{Pettini:2004fk,Pilyugin:2005,Pilyugin:2010aa,Pilyugin:2012_2,Marino:2013ty,Pilyugin:2016aa}), or theoretically, in which oxygen abundance have been inferred via photoionization models (e.g. \citealt{McGaugh:1991aa,Zaritsky:1994aa,Kewley:2002aa,Kobulnicky:2004aa,Tremonti:2004aa,Dopita:2013aa,Dopita:2016aa}), or be a combination of the two. 
Unfortunately, comparisons among metallicities estimated through different calibrations reveal large discrepancies, even for the same sample of objects, with variations up to $\sim 0.6$ dex \citep{Moustakas:2010aa, Kewley:2008aa}.
In fact, theoretical calibrations are known to produce higher metallicity estimates with respect to empirical calibrations based on the \Te\ method. The origin of these discrepancies is still unclear, but they could be attributed on one hand to oversimplified assumptions made in most of the photoionization models, e.g. about the geometry of the nebulae and the age of the ionizing source \citep{Moustakas:2010aa} and on the other hand to temperature gradients and fluctuations that may cause an overestimate of the electron temperature and a consequent underestimate of the true metallicity with the \Te\ method \citep{Peimbert:1967qv, Stasinska:2002lr, Stasinska:2005aa}.
Great care is therefore needed when using composite calibrations built with different methods over different metallicity ranges, due to the large uncertainties introduced on the absolute metallicity scale.
Empirical calibrations are generally preferable because they are based on the \Te\ method abundance scale, which is directly inferred from observed quantities. 
Moreover, on the abundance scale based on photoionization models the Milky Way, where abundances can be precisely measured, would represent a very peculiar galaxy, falling well below the M-Z defined by similar star forming galaxies. The discrepancy is reduced by more empirical metallicity calibrations that provide lower abundances.
At the same time, one of their main limitations is that they are often calibrated for samples of objects that do not properly cover all the galaxy parameters space; this means, for example, that empirical calibrations obtained from a sample of low excitation \Hii regions could give unreliable results when applied to global galaxy spectra. 
Recently, the application of integral field spectroscopy allowed to study galaxy properties in great detail and to extend the \Hii regions database for compiling abundances in order to obtain calibrations based on the \Te\ method (e.g. \citealt{Marino:2013ty} for the CALIFA survey). 
However, self-consistent calibrations obtained from integrated galaxy spectra and covering the entire metallicity range are still scarce. 

In this work we derive a set of new empirical calibrations for some of the most common strong line metallicity indicators, thanks to a uniform application of the \Te\ method over the full metallicity range covered by SDSS galaxies. 
We combined a sample of low metallicity galaxies with \oiii$\lambda 4363$ detection from the SDSS together with stacked spectra of more than $110\,000$ galaxies in bins of log[O{\sc ii}]/H$\beta$ - log[O{\sc iii}]/H$\beta$ that allowed us to detect and measure the flux of the crucial auroral lines needed for the application of the \Te\ method also at high metallicity.
Other studies demonstrated the potentiality and reliability of the stacking technique \citep{Liang:2007aa,Andrews:2013ol,Brown:2016aa}; compared  to these works, our approach differs in the sense that we do not rely on any assumption regarding the nature and the form of the relationships between metallicity, mass and SFR, but only on the hypothesis that oxygen abundance can be determined from a combination of \oii and \oiii emission line ratios.

The paper is organized as follows: in Section \ref{Sec:method} we describe the sample selection and the procedure used to stack the spectra, subtract the stellar continuum and fit the emission lines of interest.
Section \ref{sec:temp_abund} describes the method we used to derive electron temperatures and chemical abundances.
We then discuss the relations between different temperature diagnostics and between temperatures of different ionization zones.
In Section \ref{Sec:stack_tests} we report some tests we performed to verify the consistency 
of our hypothesis and stacking procedure.
In Section \ref{Sec:calib} we present our new empirical calibrations for some of the most common strong-line abundances diagnostics and we compare them with previous ones from literature.
We then apply them to the original SDSS sample as a test of their self consistency.
Section \ref{sec:summary} summarize our main results.

A publicly available user-friendly routine to apply our new calibrations can be found on the webpage \newline 
\url{http://www.arcetri.astro.it/metallicity/}.

\section{Method}
\label{Sec:method}

\subsection{Sample Selection}

Our galaxy sample come from the SDSS Data Release $7$ (DR$7$; \citealt{Abazajian:2009aa}), a survey including $\sim 930000$ galaxies in an area of $8423$ square degrees. Emission line data has been taken from the MPA/JHU\footnote{\url{http://www.mpa-garching.mpg.de/SDSS/DR7/}} catalog, in which also stellar masses \citep{Kauffmann:2003lr}, SFRs \citep{Brinchmann:2004lr,Salim:2007aa} and metallicities \citep{Tremonti:2004aa} are measured.
We chose only galaxies with redshifts in the range $0.027 < z < 0.25$, to ensure the presence of the 
\oii$\lambda 3727$ emission line and of the \oii$\lambda\lambda 7320,7330$ doublet within the useful spectral range of the SDSS spectrograph ($3800$-$\SI{9200}{\angstrom}$).
We selected only galaxies classified in the MPA/JHU as star forming, discarding galaxies dominated by AGN contribution according to criteria for BPT-diagram classification illustrated in \cite{Kauffmann:2003aa}, in order to avoid any effect on the emission line ratios that could cause spurious metallicity measurements. 
We also used a SNR threshold of $5$ on the H$\alpha$, H$\beta$, \oiii$\lambda 5007$ and \oii$\lambda 3727$  emission line fluxes.
After applying these selection criteria the total number of galaxies in our sample was reduced to $118478$, with a median redshift of $z=0.072$.
At this redshift, the $3''$ diameter of the SDSS spectroscopic fiber corresponds to $\sim 3$ kpc.

\subsection{Stacking procedure}
\begin{figure*}
\centering
\includegraphics[width=2.\columnwidth, trim = 0.5cm 0.5cm 0.5cm 0cm]{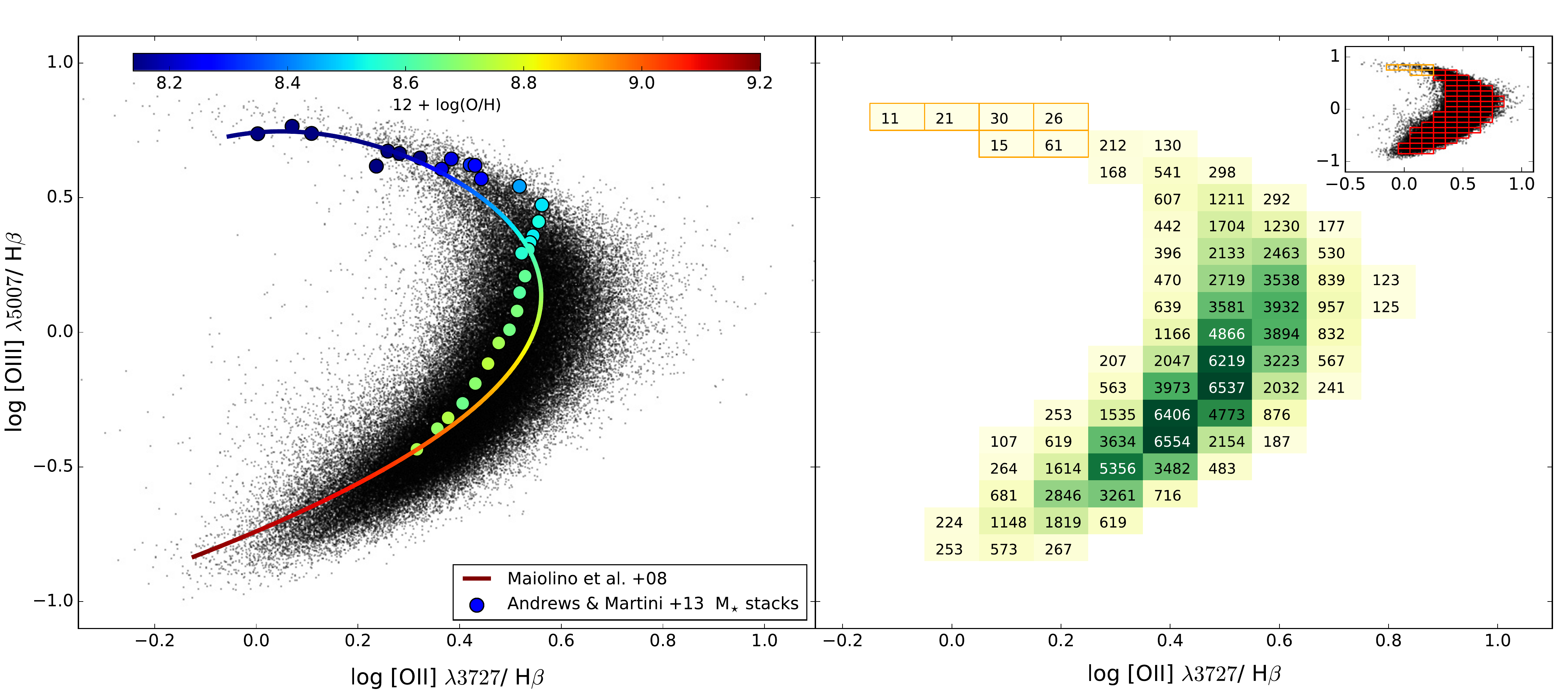}
\label{fig:mappa_ogg}
\caption{\textit{Left Panel}: The distribution of our galaxy sample in the log \oii$\lambda3727$/H$\beta$ - log \oiii$\lambda5007$/H$\beta$ diagram.
The curve represents the combined calibrations for the [O{\sc ii}]/H$\beta$ and [O{\sc iii}]/H$\beta$ metallicity indicators from \citet{Maiolino:2008cv}, color coded by the metallicity inferred from the combination of the two indicators.
The \citet{Andrews:2013ol} stacks in bins of stellar mass are shown as circle points and color coded for their direct metallicity measurement.
\textit{Right Panel}: Stacking grid for our sample of SDSS galaxies in the log \oii$\lambda3727$/H$\beta$ - log \oiii$\lambda5007$/H$\beta$ diagram.
Each square represents a $0.1$ x $0.1$ dex$^{2}$ bin, color-coded by the number of galaxies included in it, which is also written for each bin.
Orange boxes represent stacks of low metallicity galaxies for which we relaxed the $100$-object threshold in the definition of our grid.
In the upper right box of the panel our stacking grid is shown superimposed on the distribution of galaxies in the diagram.}
\end{figure*}

Our primary goal is to perform accurate measurements of galaxy metallicity in order to obtain more consistent calibrations for the main strong-line indicators, thanks to a uniform application of the \Te\ method.
Unfortunately, in distant galaxies the \oiii$\lambda 4363$ and \oii$\lambda\lambda 7320,7330$ auroral lines are too weak to be detected in the individual spectra at metallicities higher than $12 + \text{log(O/H)} \gtrsim 8.3$.
Thus, we decided to stack spectra for galaxies that are expected to have similar metallicities.

Galaxies are stacked according to their values of reddening corrected [O{\sc ii}]$\lambda 3727$/H$\beta$ and [O{\sc iii}]$\lambda 5007$/H$\beta$ flux ratios. This is based on the assumption that the so called strong-line methods can be used to discriminate the metallicities of star forming galaxies when multiple line ratios are simultaneously considered.
We stress that we are not assuming that a particular combination of these line ratios, such as R$_{23}$, is related to metallicity, but only that galaxies with simultaneously the same values of both [O{\sc iii}]$\lambda 5007$/H$\beta$ and 
[O{\sc ii}]$\lambda 5007$/H$\beta$ have approximately the same oxygen abundance.
In fact, these are the two line ratios directly proportional to the main ionization states of oxygen and are thus individually used as metallicity diagnostics \citep{Nagao:2006gd, Maiolino:2008cv}.
Moreover, their ratio [O{\sc iii}]/[O{\sc ii}] is sensitive to the ionization parameter and it 
is also used as an indicator of oxygen abundance, especially in metal enriched galaxies, 
due to the physical link between ionization and gas-phase metallicity (e.g. \citealt{Nagao:2006gd,Masters:2016aa}).
This means that the location of a galaxy on the [O{\sc ii}]$\lambda 3727$/H$\beta$-[O{\sc iii}]$\lambda 5007$/H$\beta$ diagram is primarily driven by the metal content and the ionization properties of galaxies. Since the scatter in a given line ratio at fixed metallicity is often regarded as driven by variations in the ionization parameter (e.g. \citealt{Kewley:2002aa}, \citealt{Lopez-Sanchez:2012aa}, \citealt{Blanc:2015aa}) our binning choice takes into account this possible source of scatter.

The left panel of Figure 1 shows the distribution of our selected SDSS galaxies in the log [O{\sc ii}]$\lambda 3727$/H$\beta$ - log [O{\sc iii}]$\lambda 5007$/H$\beta$ diagnostic diagram.
We overplot the semi-empirical calibration of \cite{Maiolino:2008cv} for the [O{\sc ii}]$\lambda 5007$/H$\beta$ and [O{\sc iii}]$\lambda 5007$/H$\beta$ indicators in order to better visualize how the position on the 2d-diagram given by the combination of these line ratios represent a metallicity sequence. The curve, color coded for the metallicity inferred from the combination of the two indicators, follows quite tightly the distribution of galaxies on the map, showing how metallicity increases from the upper left region of the diagram to the bottom left one.
To further illustrate how metallicity varies along this diagram we can also use the metallicity obtained with the \Te\ method from composite spectra in bins of stellar mass by \cite{Andrews:2013ol}, whose stacks are shown as circled points in the left panel of the figure.
Also in this case we can recognize a pattern in which their mass stacks, each point being representative of the line ratios measured from the associated composite spectra, increase  monotonically in metallicity following the galaxy sequence on the diagram. 
Thus, both methods reveal a clean variation of oxygen abundance with location on the diagram, though being based on different and independent approaches; this strengthens the idea of using the combination of [O{\sc ii}]/H$\beta$ and [O{\sc iii}]/H$\beta$ as a metallicity indicator.
We note that differences among metallicity values predicted by the \cite{Maiolino:2008cv} calibrations and the \cite{Andrews:2013ol} stacks (with the first ones predicting higher abundances than the latter) is only due to the different abundance scale upon which the two methods are defined, being the \cite{Maiolino:2008cv} indicators calibrated with photoionization models at high metallicities and the \cite{Andrews:2013ol} stacks based on \Te\ method metallicities.

In Sec. \ref{Sec:stack_tests} we test our assumptions by comparing \Te\ metallicities inferred from single galaxy spectra belonging to the same bin; this allows also to evaluate the main issues related to the stacking technique (see also the discussion in Sec. \ref{Sec:trel}).
We refer to these sections for an exhaustive discussion on this topic.

We thus created stacked spectra in bins of $0.1$ dex of log [O{\sc ii}]/H$\beta$ and log [O{\sc iii}]/H$\beta$.
The choice of the $0.1$ dex width in the binning grid represents a good compromise between keeping an high enough number of galaxy in each bin to ensure auroral line detection and at the same time avoid wider bins in which we could have mixed object with too different properties.
We performed some tests in stacking spectra and computing oxygen abundance with different bin sizes, finding no relevant differences.
\begin{figure*}	
\centering
\includegraphics[width=2.\columnwidth, trim=0.7cm 1.5cm 0.7cm 0cm]{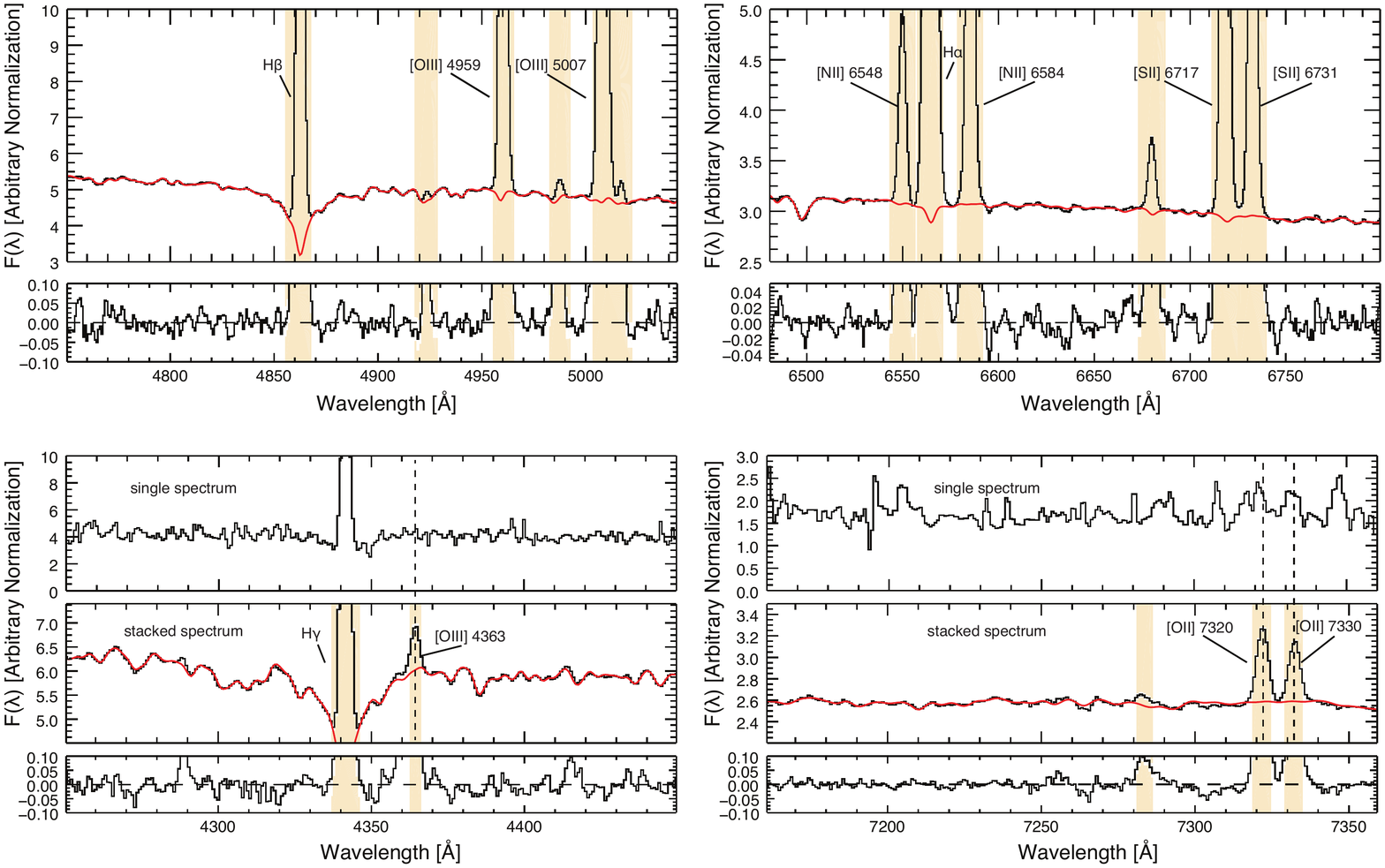}
\caption{Fit and subtracted spectra for wavelength ranges relative to H$\beta$ and \oiii nebular lines (upper left panels),  
H$\alpha$, \nii and \sii nebular lines (upper right panels), \oiii$\lambda 4363$ auroral line 
(lower left panels) and \oii$\lambda 7320,7330$ auroral lines (lower right panels) respectively, for the $0.5;0.5$ stack. For strong nebular lines, the upper panel shows the stacked spectrum (black) and the stellar continuum best-fit component (red), while the bottom panel shows the residual spectrum after the stellar continuum subtraction. For auroral lines boxes, a single galaxy spectrum is shown in the upper panel for comparison, while the stacked spectrum is shown in the middle one. The yellow shaded regions mark the spectral interval masked out during the stellar continuum fitting procedure.}
\label{fig:stacked_spectra}
\end{figure*}

We adopted the emission line values provided by the MPA/JHU catalog to create the set of galaxy stacks.
All line fluxes have been corrected for Galactic reddening, adopting the extinction law from \cite{Cardelli:1989lr} and assuming an intrinsic ratio for the Balmer lines H$\alpha$/H$\beta = 2.86$ (as set by case B recombination theory for typical nebular temperatures of $T_{e}=$ $\SI{10000}{\kelvin}$ and densities of $n_{e}\approx$ $\SI{100}{\cm}^{-3}$).

In the right panel of Figure \ref{fig:mappa_ogg} the binning grid for our galaxy sample in the space defined by log(\oii$\lambda 3727$/H$\beta$) and log(\oiii$\lambda 5007$/H$\beta$) is shown, color coded by the number of objects in each bin.
In the up-right corner of the figure we show the distribution of the galaxy sample in the diagnostic diagram, with our binning grid superimposed. 
In the construction of our binning grid we required a minimum of $100$ sources per bin: this was a conservative choice in order to average  enough galaxy spectra to ensure the required SNR (i.e. at least $3$) on auroral lines detection after the stacking procedure.  
Since we imposed a threshold of 100 sources per bin, the upper-left corner of the diagram, occupied by the galaxies of lower metallicity in the sample, is not well covered by our stacking grid. For this reason, we extended our grid to include also low-metallicity galaxies by reducing the threshold to $10$ sources in that area of the diagram, enough to detect auroral lines in stacked spectra with a SNR higher than $3$ in this metallicity regime.
This extension of the grid is marked with orange borders in the figure. This allows our grid to entirely cover the region occupied by SDSS galaxies, probing the largest possible combination of physical parameters in the sample. 
Throughout this paper we will refer to a particular stack by indicating the center of the corresponding bin in both the line ratios considered (e.g. $0.5;0.2$ corresponds to the bin centered in log\oii$\lambda 3727$/H$\beta=0.5$ and log\oiii$\lambda 5007$/H$\beta=0.2$).

Before creating the composite spectrum from galaxies belonging to the same bin, each individual spectrum has been corrected for reddening with a \cite{Cardelli:1989lr} extinction law and normalized to the extinction corrected H$\beta$. We have verified that the final results do not depend on the choice of the extinction law, by alternatively using the \cite{Calzetti:1994aa} extinction law in a few random bins.
Then, each spectrum has been re-mapped onto a linear grid ($3000$-$\SI{9200}{\angstrom}$), with wavelength steps of $\Delta \lambda = \SI{0.8}{\angstrom}$, and shifted at the same time to the rest frame to compensate for the intrinsic redshift of the sources. 
This procedure may cause a redistribution of the flux contained in a single input channel to more than one output channel; in order to take into account this effect, the incoming flux is weighted on the overlap area between the input and output channels.
Finally, to create the stacked spectra we took the mean pixel by pixel between the $25^{\text{th}}$ and the $75^{\text{th}}$ percentile of the flux distribution in each wavelength bin; in this way we could avoid biases introduced by the flux distribution asymmetry clearly visible in every flux channel as a right-end tail.

\subsection{Stellar continuum subtraction}

Stacking the spectra improve significantly the SNR of the auroral lines, but we must also fit and subtract the stellar continuum to accurately measure their fluxes.
To perform the stellar continuum fit and subtraction on our stacked spectra we have created a synthetic spectrum using the MIUSCAT library of spectral templates \citep{Vazdekis:2012aa,Ricciardelli:2012aa}, an extension of the previous MILES library 
\citep{Falcon-Barroso:2011qy,Cenarro:2001uq,Vazdekis:2010lr,Sanchez-Blazquez:2006fk} in which both 
Indo-U.S. and CaT libraries have been added to fill the gaps in wavelength coverage.
The new MIUSCAT library covers a wavelength range of $3525$-$\SI{9469}{\angstrom}$, although the useful spectral window for this work is entirely covered by MILES templates, whose resolution is $\SI{2.51}{\angstrom}$(FWHM).
Stellar templates have been retrieved from the MILES website\footnote{\url{http://miles.iac.es}} for a 
wide range of ages and metallicities, assuming an unimodal initial mass function with a $1.3$ slope (i.e. a Salpeter IMF).
The stellar continuum subtraction in the \sii $\lambda 4069$ spectral window (close to H$\delta$) has been performed using a different kind of stellar templates, the P\'EGASE HR\footnote{\url{http://www2.iap.fr/pegase/}}\citep{Le-Borgne:2004aa}, a library which covers a wavelength range of $4000$-$\SI{6800}{\angstrom}$ with a spectral resolution of $R=10000$ at $\lambda=\SI{5500}{\angstrom}$;
this allowed a better stellar continuum fit in the proximity of the ~\sii $\lambda 4069$ auroral line.
To further improve emission line fluxes measurements, stellar continuum fits and subtractions have been performed selecting subregions of the spectrum centred on the lines of interest, each subregion being large a few hundred angstrom. 
During the procedure the location of the emission lines have been masked out in order to prevent the fit to be affected by non stellar features. 
We performed the fit exploiting the IDL version of the penalized pixel-fitting (pPXF) procedure by \cite{Cappellari:2004kx}.
In Table \ref{tab:intervalli_maschere} are reported, for each emission lines whose flux have been measured in this work, the spectral window of the stellar continuum fit and the wavelength range that has been masked out.

\begin{table}
\caption[Spectral windows and Mask ranges of measured emission lines.]{(1) Emission lines. (2) Wavelength range of stellar continuum fit. 
(3) Spectral range that was masked out.}
\label{tab:intervalli_maschere}
\centering
\begin{tabular}{ccc}
\toprule
Line 													& Fit Range	 						& Mask Range		  \\
														& [\AA] 								& [\AA]  			  		\\
	(1)												&	(2)									&  (3)     			  		\\				
\midrule
\oii $\lambda 3727$				& $3650$-$3830$			 	&    $3723.36$-$3733.60$  \\
$[$Ne{\sc iii}$] \lambda 3870$				& $3850$-$4150$			 	&    $3866.29$-$3874.03$  \\
$[$S{\sc ii}$] \lambda 4069$					& $4000$-$4150$			 	&    $4068.39$-$4071.11$  \\
H$\delta$  $\lambda 4102$			& $3850$-$4150$			 	&    $4098.79$-$4107.00$  \\
H$\gamma$  $\lambda 4340$		& $4250$-$4450$			 	&    $4337.34$-$4346.03$  \\
\oiii $\lambda 4363$				& $4250$-$4450$			 	&    $4362.98$-$4365.89$  \\
H$\beta$  $\lambda 4861$			& $4750$-$5050$			 	&    $4857.82$-$4867.55$  \\
\oiii $\lambda 4960$				& $4750$-$5050$			 	&    $4955.33$-$4965.26$  \\
\oiii $\lambda 5007$				& $4750$-$5050$			 	&    $5003.23$-$5013.25$  \\
\nii $\lambda 5756$				& $5650$-$5850$			 	&    $5754.32$-$5758.16$  \\
\nii $\lambda 6549$				& $6480$-$6800$			 	&    $6543.30$-$6556.40$  \\
H$\alpha$  $\lambda 6563$			& $6480$-$6800$			 	&    $6558.05$-$6571.17$  \\
\nii $\lambda 6584$				& $6480$-$6800$			 	&    $6578.69$-$6591.87$  \\
\sii $\lambda 6717$					& $6480$-$6800$			 	&    $6711.57$-$6725.01$  \\
\sii $\lambda 6731$					& $6480$-$6800$			 	&    $6725.94$-$6739.40$  \\
\oii $\lambda 7320$				& $7160$-$7360$			 	&    $7318.50$-$7323.28$  \\
\oii $\lambda 7330$				& $7160$-$7360$			 	&    $7329.24$-$7334.12$  \\
\bottomrule
\end{tabular}
\end{table}
Figure \ref{fig:stacked_spectra} shows examples of the results of the stacking procedure and stellar continuum subtraction for the $0.5;0.5$ bin, in particular in spectral windows including H$\beta$, \oiii, H$\alpha$, \nii  and \sii nebular lines and \oiii$\lambda 4363$, \oii$\lambda 7320,7330$ auroral lines respectively. For latter emission lines, a single galaxy spectrum from the same stack is shown for comparison, to underline the dramatic increase in SNR that allow to reveal the otherwise invisible auroral lines. The orange regions mark the spectral range masked out from the stellar fitting around nebular lines. In each plot, the lower panel shows the residual spectrum of the fit.

\subsection{Line Flux Measurement and Iron contamination of \oiii $\lambda 4363$ auroral line at high metallicity}

We fit emission lines with a single Gaussian profile, fixing velocities and widths of the weak auroral lines by linking them to the strongest line of the same spectral region.
For doublets, we fixed the velocity width of the weaker lines to the stronger ones (\oii$\lambda 3727$ 
to \oii $\lambda 3729$, \oiii $\lambda 4960$ to \oiii $\lambda 5007$,
\nii $\lambda 6548$  to \nii $\lambda 6583$, \sii $\lambda 6731$ to \sii $\lambda 6717$ and \oii $\lambda 7330$ to \oii $\lambda 7320$).

\begin{figure}
\centering
\includegraphics[width=1.\columnwidth, trim=7.8cm .8cm 7.cm 0cm]{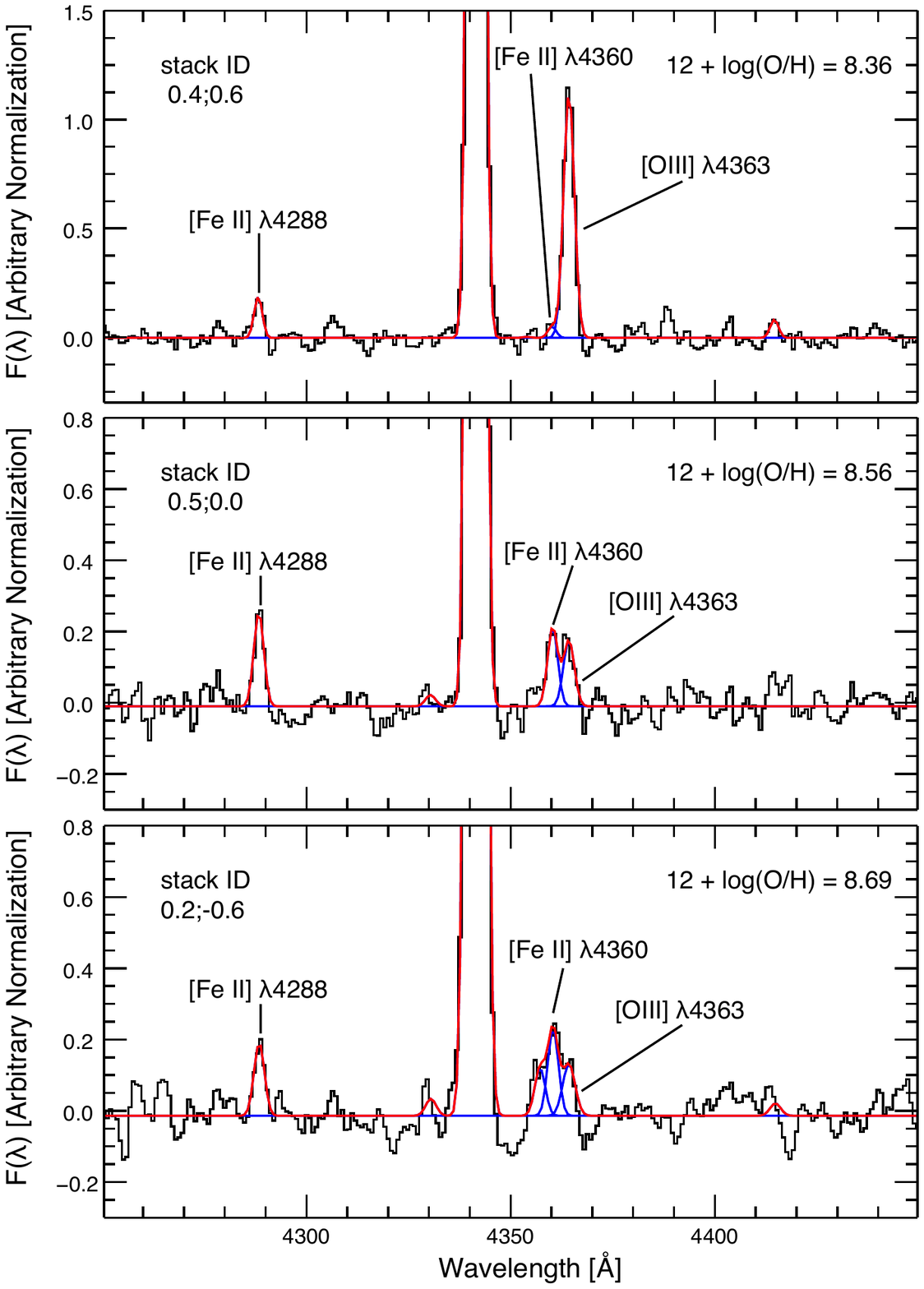}
\caption{\textit{Left}: Composite spectra for the $0.4;0.6$ (upper panel), $0.5;0.0$ (middle panel) and $0.2;-0.6$ (lower panel) stack, in the wavelength range relative to \oiii $\lambda 4363$, after the stellar continuum subtraction. The different components of the fit are reported in blue while the red curve represents the total fit. 
The metallicity of each stack is reported in the right-upper part of the corresponding panel.
The contamination of the \oiii$\lambda4363$ line becomes more relevant with increasing metallicity (in the last case we can fit up to three components), as well as the intensity of the [Fe {\sc ii}] emission line at $\SI{4288}{\angstrom}$.
}
\label{fig:4363_flag}
\end{figure}

During the fitting procedure an emission feature close to $\SI{4360}{\angstrom}$ has been detected and became blended with the \oiii $\lambda 4363$ auroral line, especially in the high metallicity stacks. A similar contamination was previously found also by \cite{Andrews:2013ol} in their composite spectra. 
The nature of this feature is unknown, but it may reasonably be associated to emission lines from [Fe {\sc ii}] $\lambda 4360$. In fact, many others features from the same ion are clearly observable both in the same (e.g. [Fe {\sc ii}] $\lambda 4288$) and in different spectral windows; this particular emission have been reported also in studies on the Orion nebula (see e.g. Table 2 of \cite{Esteban:2004kx}). 
Moreover, the strength of of the line increases with increasing metallicity, as well as the other [Fe {\sc ii}] lines in the spectra.
In Figure \ref{fig:4363_flag} we show three stacked spectra corresponding to different metallicities, namely $0.4;0.6$, $0.5;0.0$ and $0.2;-0.6$, after performing the stellar continuum subtraction in the spectral window that contains the \oiii $\lambda 4363$ line. The metallicity of each stack (see Sect. \ref{Sec:calib}) is reported on every panel.
The figure clearly shows how \oiii $\lambda 4363$ becomes more contaminated as the metallicity increases, with the [Fe {\sc ii}] emission being just a few percent of the flux of the oxygen one in the upper panel and then completely blending with it in the other two.
The [Fe {\sc ii}] emission line at $\SI{4288}{\angstrom}$ is also clearly visible in all the composite spectra, with increasing strength for increasing metallicity, as expected.

Therefore we simultaneously fit the $\lambda 4360$ feature and \oiii $\lambda 4363$, linking both velocity widths and central wavelengths to H$\gamma$. The different components of the fit are shown in blue in Figure \ref{fig:4363_flag}, with the red line representing the total fit.
Similarly to \cite{Andrews:2013ol}, we consider the fit not sufficiently robust when the $\lambda 4360$ emission flux resulted $\geq 0.5$ times the flux measured for \oiii $\lambda 4363$. 
The use of contaminated \oiii$\lambda4363$ line would result in totally non-physical temperatures, which result overestimated by a factor of ten. 
According to this criteria, $42$ out of $69$ bins have been flagged for undetected \oiii $\lambda 4363$. 
We note that many previous detection of the \oiii $\lambda 4363$ may be possibly contaminated by this feature, resulting in unreliable measurements for this crucial auroral line; therefore we recommend great care in using \oiii $\lambda 4363$, when detected, to measure electron temperature from high metallicity (12+log(O/H)$\geq 8.3$) galaxy spectra.
In the next section we will discuss how to derive the electron temperature for the high ionization zone for those stacks where the \oiii $\lambda 4363$ was not considered sufficiently robust.

Despite the large number of galaxies in each bin and the great care in the fitting procedure, in some of our stacks we were unable to measure both \oii and \oiii auroral line fluxes with sufficient precision.
These are the stacked spectra corresponding to the $0.0;-0.8$, $0.1;-0.4$, $0.1;-0.7$, $0.1;-0.8$, $0.8;0.2$ and $0.6;-0.3$ bins and we decided to exclude this stacks from all the forthcoming analysis.

\section{Electron Temperatures and Ionic Abundances Determination}
\label{sec:temp_abund}

\subsection{Electron Temperatures}

In principle, to measure electron temperatures and densities of different zones, the complete ionization structure of a \Hii region is needed. This is actually not possible and simpler approximations are always used. Usually a two-zone (of low and high ionization) or even a three-zone (of low, intermediate and high ionization) structure is adopted to model the \Hii regions responsible for emission lines in galaxies. In this work we consider a two-zone \Hii region:
in this scenario, the high ionization zone is traced by the O$^{++}$ ion, while the low ionization zone could be traced by different ionic species, e.g. O$^{+}$, N$^{+}$ and S$^{+}$.
Thus, given the SDSS spectral coverage, in our case we have three different diagnostics for the temperature of the low ionization zone (which we will refer to as $t_{2}$ from now on), namely \oii $\lambda 3727,3729$/\oii $\lambda 7320,7330$, \nii$\lambda 6584$/\nii$\lambda 5755$ and \sii$\lambda 6717,6731$/\sii$\lambda 4969$, but only one for the temperature of the high ionization zone ($t_{3}$), namely \oiii$\lambda 5007$/\oiii$\lambda 4363$.
Other collisionally excited lines probing the temperature of the intermediate and high ionization region are either too weak and thus undetectable even in galaxy stacks (e.g. [Ar {\sc iii}] $\lambda 5192$) or fall outside the spectral range of the SDSS spectrograph (e.g. [S {\sc iii}] $\lambda 9069$, [Ne {\sc iii}] $\lambda 3342$) and we could not use them.

We computed electron temperatures exploiting PyNeb \citep{Luridiana:2012aa, Luridiana:2015aa}, the Python-based version of the \emph{stsdas nebular} routines in IRAF, using the new atomic dataset presented in \cite{Palay:2012aa}.
This routines, which are based on the solution of a five level atomic structure following \cite{De-Robertis:1987rm}, determine the electron temperature of a given ionized state from the nebular to auroral flux ratio assuming a value for electron density.
The electron density n$_\text{{e}}$ can be measured from the density sensitive \sii $\lambda\lambda 6717,6731$ or \oii $\lambda\lambda 3727,3729$ doublets.
In the majority of our stacks we fall in the low density regime (n$_\text{{e}}$ < $\SI{100}{cm^{-3}}$), for we measure for example a \sii ratio close to the theoretical limit of $1.41$; in this cases the dependence of our temperature diagnostics upon density is small.
We note that using older atomic datasets (e.g. \citealt{Aggarwal:1999aa}), instead on the new ones by \cite{Palay:2012aa}, would result in similar t$_{2}$ but t$_{3}$ systematically 
higher on average by $400$ K. 
This is consistent with expectations given the updated effective collision strengths for \oiii lines, as pointed out e.g. in \cite{Nicholls:2013aa} where a discrepancy of $\sim 500\ \text{K}$ is expected at T\oiii $\sim 10^{4}\ \text{K}$ (see e.g. Section 7 and Figure 2 and 12 of their paper for further details).
We also note that the collision strengths presented in \cite{Palay:2012aa} for the \oiii optical transitions are tabulated for a wide range of temperatures typical of nebular environments (from $100$ K to $30000$ K), which include all the temperatures we expect to 
find given the metallicity range spanned by the SDSS galaxies.
Temperatures uncertainties were computed with Monte Carlo simulations.
We generated $1000$ realizations of the flux ratios, following a normal distribution with $\sigma$ equal to the errors associated to the flux measurement by the fitting procedure and propagated analytically, and for each of them a temperature value was calculated. 
Then, we took the standard deviation of the resulting distribution as the error to associate to our temperature measure.
\begin{figure*}
\centering
\includegraphics[width=2.\columnwidth, trim = 1.5cm 0.6cm 1.5cm 0cm]{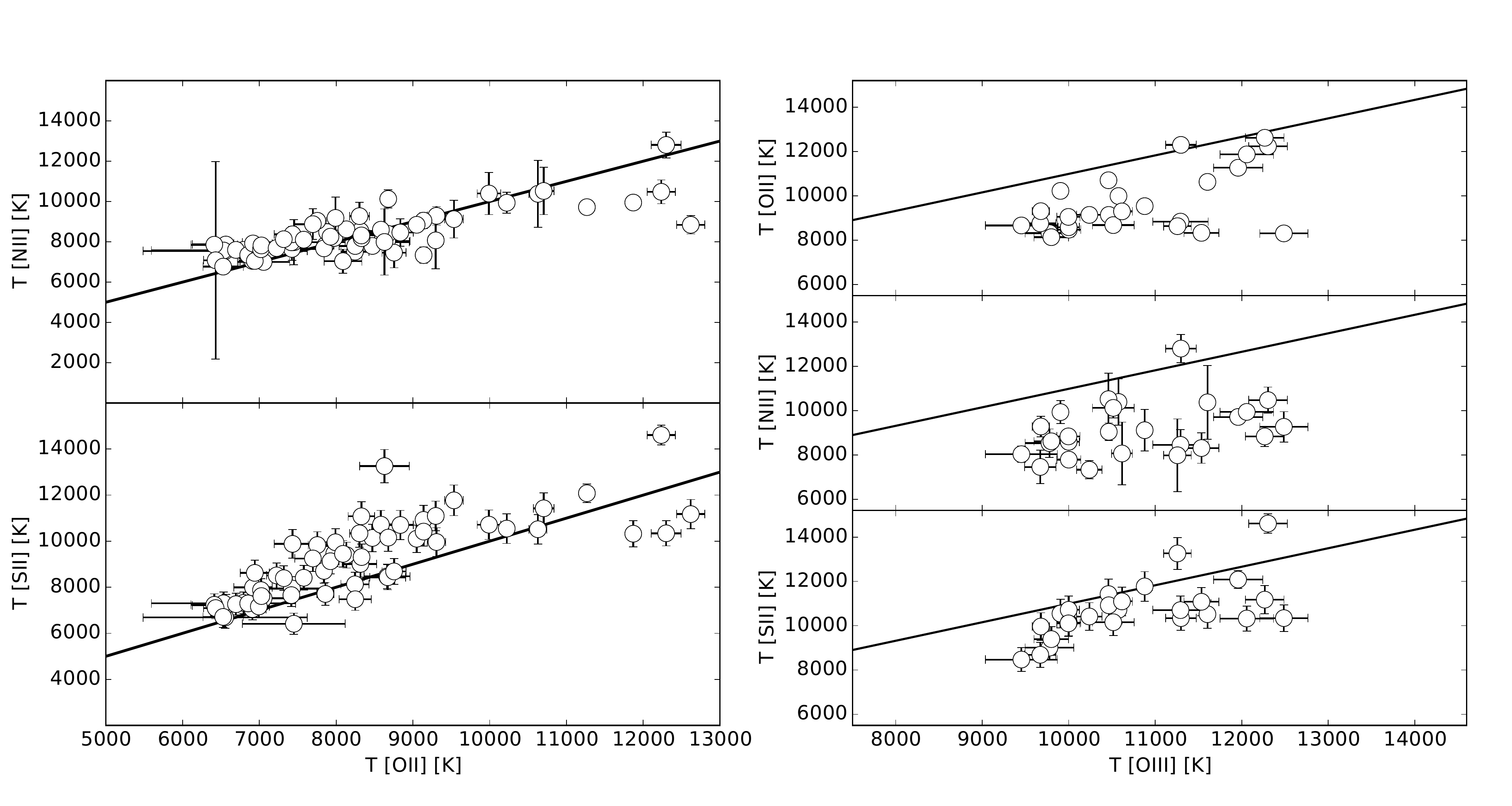}
\caption[t2_t3_plots]{\textit{Left Panels}: Electron temperatures derived from the \nii \ and \sii \ line ratios as a function of the electron temperature derived from \oii; the equality line is shown in black. While the \nii temperatures are consistent with the \oii ones, the \sii provides temperatures systematically higher. \textit{Right Panels}: Electron temperatures of the low ionization zone derived with all three different diagnostics as a function of the electron temperature of the high ionization zone derived from \oiii line ratio. The black line represents the \trel from equation \ref{eq:t2-t3}, which does not provide a good representation of the data.}
\label{fig:t2_t3_diagrams}
\end{figure*}

The left panels of Figure \ref{fig:t2_t3_diagrams} show the relations between the temperatures of the low ionization zone of our stacks inferred through different diagnostics, i.e. \Te \nii (upper panel) and \Te \sii (lower panel) as a function of \Te \oii; the black line represents the line of equality. 
In the upper panel, we can see how the electron temperatures derived from nitrogen line ratios are consistent with \Te \oii, although with a large scatter, while in the lower panel we show that \Te \sii is larger than \Te \oii for almost all of our points, thus overpredicting t$_{2}$ with respect to that derived through oxygen lines.
Evidences of similar temperatures discrepancies have been reported by several works in the literature aimed at studying the physical properties of single \Hii regions (see e.g. \citealt{Kennicutt:2003fk,Bresolin:2005aa,Esteban:2009gu,Pilyugin:2009aa,Binette:2012aa,Berg:2015aa}).
Interestingly, when \Te \sii and \Te \oii are considered in these papers, average offsets are usually found in the direction of larger \Te \oii, differently from what we found for our stacks. 
The most likely explanation resides  in the different atomic dataset for energy levels and collision strengths used among these works and ours.
In fact, when computing \Te \sii for our stacks exploiting different datasets, we find variations up to thousands kelvins even at fixed diagnostic ratio.

Temperature fluctuations and inhomogeneities as well as shocks propagating within the photoionized gas  have been proposed as the main sources of discrepancies between \Te\ inferred through different ionic tracers. 
Moreover, we are here considering the simple case of Maxwell-Boltzmann distributed electrons, while recent studies suggested that k-distributions could better represent the behavior of free electrons inside single \Hii regions \citep{Nicholls:2012aa,Nicholls:2013aa,Dopita:2013aa}.
In particular, when considering only Maxwell-Boltzmann distributions, the electron temperature inferred using the most common diagnostics could be overestimated and this effect is more relevant for ions in which the excitation temperature of the upper level involved in the transitions results different from the kinetic temperature of the distribution \citep{Nicholls:2012aa}: the effect of k-distributed electrons could therefore affect different temperature diagnostics in different ways and thus explain the observed discrepancies in \Te\ estimates. 
For an in-depth discussion on how k-distributed electrons could affect the main metallicity diagnostics in \Hii regions, see \cite{Dopita:2013aa}.
However, it is not clear to what extent these processes can affect the determination of electron temperature when considering global galaxy spectra and, in particular, a stacking of many galaxies, as we do in this work.

\subsection{The \trel}
\label{Sec:trel}
In the right panels of Figure \ref{fig:t2_t3_diagrams} the relations between the temperatures of the low and high ionization zones are shown for those stacks in which we have been able determine \Te \oiii directly from the spectra.
Many works in literature report the existence of a relation between the temperatures of the different ionization zones. 
The linear form of this relation, called \trel, have been proposed for the first time by \cite{Campbell:1986lr} and then revised in several studies 
\citep{Garnett:1992aa,Izotov:2006aa,Pilyugin:2006aa,Pilyugin:2007,Pilyugin:2009aa,Pilyugin:2010ab}. This relation is of great interest in the context of nebular studies since it is generally used to compute the electron temperature for unseen ionization states.
In this work we will consider the linear relation suggested by \cite{Pilyugin:2009aa} for temperatures derived through oxygen lines in their sample of \Hii regions, given by the equation
\begin{equation}
t_{2} = 0.264 + 0.835 \ t_{3}\ ,
\label{eq:t2-t3}
\end{equation}
where both temperatures are in units of ${10}^{4}$K. 
The \trel of equation \ref{eq:t2-t3} is shown for comparison in each plot as a black line.
We can see that the great majority of our stacks falls below the \trel independently of the type of ion tracer, although with different median 
offsets from the relation corresponding to different tracers.
In any case, it is clear that this \trel underestimates the temperature of the high ionization zone (or overestimates the temperature of the low ionization zone) for our composite spectra.
A very similar result was found by \cite{Andrews:2013ol} for their stacked spectra in bins of stellar mass, even though they used a different, but quite similar, form for the \trel.
The median offsets from the \trel for our stacks are  $-2642$ K, $-2712$ K and $-1055$ K for \Te\oii, \Te\nii and \Te\sii respectively.
\begin{figure*}
\centering
\includegraphics[width=2.\columnwidth, trim = 1.5cm 0.cm 1.5cm .5cm]{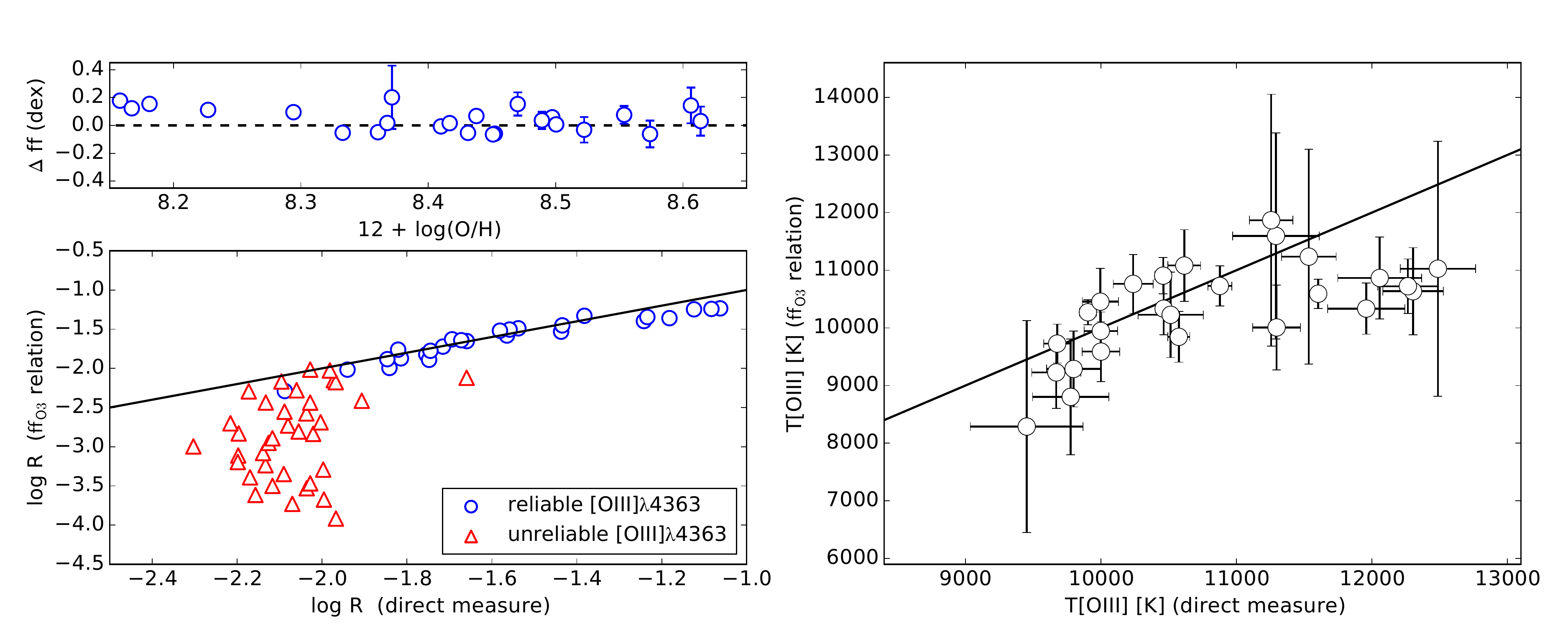}
\caption{\textit{Left Panel} : Log R (i.e. log \oiii$\lambda 4363$/H$\beta$) directly measured from the stacked spectra as a function of the same quantity obtained through the ff relation. Blue circles represent stacks whose \oiii$\lambda 4363$ detection was considered robust according to the criteria described in the text, while red triangles are stacks whose \oiii$\lambda 4363$ detection was considered unreliable. The black line represent the ff$_{\text{O}3}$ relation of equation \ref{eq:ff_rel}.
In the upper box the offset of log(R) from the ff$_{\text{O}3}$ relation for the stacks with reliable \oiii$\lambda 4363$ measurements is plotted as a function of the metallicity of the stacks.
\textit{Right Panel} : \Te \oiii derived from the ff$_{\text{O}3}$ relation as a function of direct measure \Te \oiii for the stacks with detected \oiii$\lambda 4363$; black line represents equality.}
\label{fig:ff_rel}
\end{figure*}

The offset between the electron temperatures of the stacks and the \trel is in agreement with the trend found by \cite{Andrews:2013ol} for galaxy stacks but also by \cite{Pilyugin:2010ab} for individual galaxies, suggesting that this effect is not a product of the stacking procedure but rather reflects the intrinsic properties of global galaxy spectra.  
The most likely explanation indeed is that galaxy spectra are the result of several contributions from \Hii regions that could present very different physical properties, both in terms of chemical composition and hardness of their ionizing sources: this may affect the auroral line fluxes in the sense that they are weighted differently in \Hii regions of different temperatures. Since the auroral line flux does not scale linearly with metallicity, the effect of a luminosity-weighted average towards warmer \Hii regions on their total flux can be substantial and difficult to account for, in a way similar to temperature fluctuations for single \Hii regions described by \cite{Peimbert:1967qv}; therefore one can obtain results that do not agree with the observed \trel for single \Hii regions \citep{Kobulnicky:1999qy,Kennicutt:2003fk}. For example, \cite{Pilyugin:2010ab} showed that the \trel offset can be substantially reproduced considering composite spectra obtained mixing contributions from few \Hii regions of very different temperatures.
Moreover, the variation of the relative contribution of each \Hii region for different ionic species, together with the contribution from diffuse ionized gas \citep{Moustakas:2006aa}, can explain the different distributions in the t2-t3 plane for different temperature diagnostics as well as the offset between different estimations of the temperature of the low ionization zone. 
Despite these difficulties, the \trel has been widely used in literature to compute electron temperatures of unseen ionization states.

\subsection{The ff relations}
Another possibility to solve the problem of determining the temperature of the high ionization zone for stacks with unmeasured \oiii $\lambda 4363$ is to rely on the relationship between the strong emission lines and the auroral line itself.
As pointed out by \cite{Pilyugin:2005aa}, a relation between auroral and nebular oxygen line fluxes has been demonstrated for \Hii regions of metallicity higher than 12 + log(O/H) $\sim 8.25$. This so called "ff relation" allow to estimate the auroral line flux from the measured nebular line fluxes when the first is not available. In this work we employed the following ff relation (which we will refer to from now on as the ff$_{\text{O}3}$ relation) obtained by \cite{Pilyugin:2006ab} to infer the \oiii $\lambda 4363$ flux for stacks where the flux of this line was not properly measured : 
\begin{equation}
\text{log R} = -4.151 - 3.118\  \text{log P} + 2.958\  \text{log R}_{3} - 0.680\  \text{(log P)}^{2}
\label{eq:ff_rel}
\end{equation}
where $\text{R}=\text{I}_{\text{\oiii}\lambda4363}/\text{I}_{\text{H}\beta}$, $\text{R}_{3} = \text{I}_{\text{\oiii}\lambda4949,5007}/\text{I}_{\text{H}\beta}$ and
P(i.e. the excitation parameter) = $\text{R}_{3}/(\text{R}_{3}+\text{R}_{2}$), with $\text{R}_{2}=\text{I}_{\text{\oii}\lambda3727,3729}/\text{I}_{\text{H}\beta}$.

Inspection of the left panel of Figure \ref{fig:ff_rel}, where log(R) obtained through the ff$_{\text{O}3}$ relation is plotted against its direct measure from the spectra, reveals that our stacks which satisfy the criteria for good \oiii$\lambda 4363$ detection (blue circle points) are in good agreement with equation \ref{eq:ff_rel}, represented by the black line.
We also plot as red triangles the points representing the composite spectra whose \oiii $\lambda 4363$ detection was flagged as unreliable due to [Fe {\sc ii}] contamination. 
Almost all of these points does not follow the ff$_{\text{O}3}$ relation, falling well below the black line of Figure \ref{fig:ff_rel}. This was expected and corroborates the fact that the \oiii $\lambda 4363$ flux measurements in those stacks can not be considered reliable.
\begin{figure*}
\centering
\includegraphics[width=2.\columnwidth,trim= 1cm 1cm 1cm 0cm]{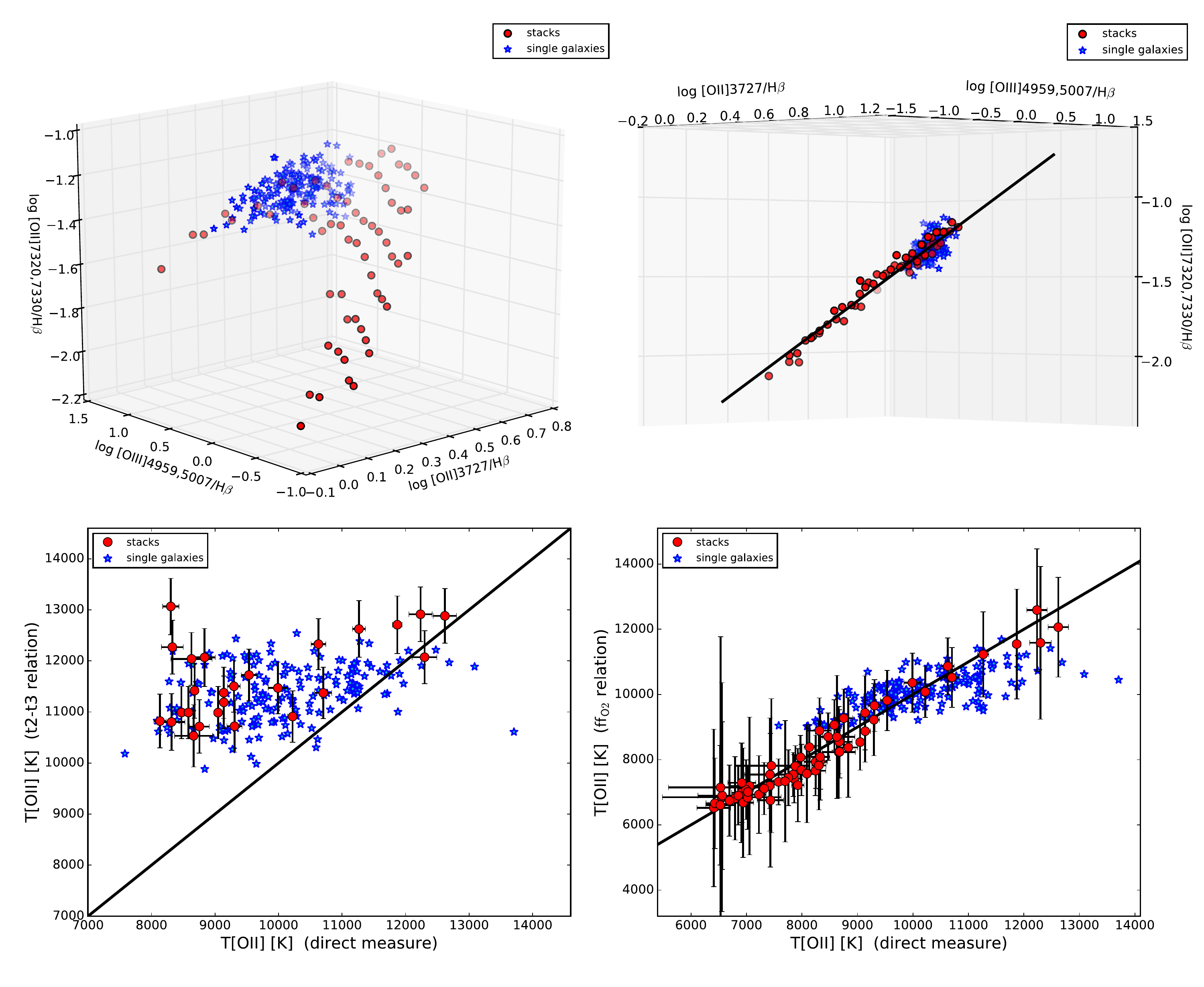}
\caption{\textit{Upper Panels} : Log \oii$\lambda 7320,7330$/H$\beta$ as a function of log \oii$\lambda 3727$/H$\beta$ and log \oiii$\lambda 5007$/H$\beta$ for the sample of our stacks (red circles) and the \citet{Pilyugin:2010ab} galaxies (blue stars). All the points lie on a tight surface and in the right panel we show the 2-D projection that minimizes the scatter and predicts the flux of the oxygen auroral doublet from a combination of the two strong line ratios; the black line represents the linear fit which defines our new ff relation.  
\textit{Bottom Panels} : \Te\oii inferred through the \trel (left panel) and through the ff$_{\text{O}3}$ relation (right panel) as a function of the direct measure \Te\oii. Symbols are the same as in the upper panels. The equality line is shown in black in both panels.}
\label{fig:ff_o2_3d}
\end{figure*}
In the upper box of the same figure the deviations of log(R) from the ff$_{\text{O}3}$ relation (defined as $\Delta \text{ff}=\text{log(R)}_{\text{direct}}-\text{log(R)}_{\text{ff relation}}$) for the stacks with good \oiii$\lambda 4363$ detection are plotted as a function of metallicity derived with the \Te\ method; the points scatter around zero with a $\sigma = 0.09$ dex, showing no trends with metallicity. 
The error bars in the upper box of Figure \ref{fig:ff_rel} represent the uncertainties on $\Delta$ff, derived propagating the errors on the line flux measurements through the equation \ref{eq:ff_rel}; with the exception of a few points, this source of uncertainty ($0.05$ dex on average) can not account for the total dispersion shown, being the larger part due to the intrinsic dispersion of the ff$_{\text{O}3}$ relation itself.

In the right panel of Figure \ref{fig:ff_rel} we compare \Te\oiii derived through the ff$_{\text{O}3}$ relation with the one directly measured from the spectra, for stacks with good detection of \oiii$\lambda 4363$; black line represents equality in this plot. 
Temperatures predicted by the ff$_{\text{O}3}$ relation are in good agreement with direct measurements within the uncertainties, and there is no evident and systematic trend unlike what happens with the \trel (see, for comparison, the upper right panel of Figure \ref{fig:t2_t3_diagrams}). 
From the above considerations and since all of our stacks, given the construction of our stacking grid, have a well defined value for R$_{2}$ and R$_{3}$ (and thus P), we decided to use the equation \ref{eq:ff_rel} instead of the \trel to determine the flux of \oiii $\lambda 4363$, and consequently the t$_{3}$, for stacks with no reliable detection of this auroral line.
This allow to minimize the systematic offset introduced on abundances determination (see also Section \ref{sec:ionic}).

\subsection{Defining an ff relation for \oii auroral lines}
Following the same idea of \cite{Pilyugin:2005aa}, we can exploit the direct measurements of \oii$\lambda 7320,7330$ in our stacks to define an analogous ff relation for the \oii auroral doublet, which we will refer to as the ff$_{\text{O}2}$ relation.
\cite{Pilyugin:2009aa} manage to obtain a similar relation 
for their sample of single \Hii regions in the low-R$_{3}$ range (i.e. log R$_{3}<0.5$).
In particular, here we search for a combination of [O{\sc ii}]/H$\beta$ and [O{\sc iii}]/H$\beta$\ (which define our stacking grid) that predicts the flux of the \oii$\lambda 7320,7330$ auroral doublet.
Inspection of the upper panels of Figure \ref{fig:ff_o2_3d} reveals that since our stacks appear to lie on a surface in the 3-D space defined by log\oii$\lambda3727$/H$\beta$ - log\oiii$\lambda5007$/H$\beta$ - log\oii$\lambda 7320,7330$/H$\beta$, we can search for the projection that minimizes the scatter
in our sample and gives the combination of the first two indices that predicts the value of the latter; such a combination could be easily formalized with a linear fit, which we show as a black line in the upper right panel of the same figure.
In order to better constrain the definition of our new ff$_{\text{O}2}$ relation, we included the sample of low metallicity SDSS DR6 galaxies from \citet{Pilyugin:2010ab} with detected \oii$\lambda 7320,7330$; these objects lie in the upper left zone of our original diagram, which is characterized by high excitation galaxies.
Even though these objects are characterized by a larger scatter than the stacks, they do not show any extreme offset from the surface defined by the stacks in the 3-D space.

The functional form of our linear fit is the following :
\begin{equation}
\text{log}\ R_{\text{\oii}} = -1.913 + 0.806\ \text{log}\ R_{2} + 0.374\ \text{log}\ R_{3}
\label{eq:my_ff}
\end{equation}
where $\text{R}_{\text{\oii}}=\text{I}_{\text{\oii}\lambda 7320,7330}/\text{I}_{\text{H}\beta}$; the results of the fit is shown in the upper right panel of Figure \ref{fig:ff_o2_3d} as the black line.
The dispersion around the ff$_{\text{O}2}$ relation is $0.04$ dex for the stacks and $0.06$ dex for  individual galaxies.

We can test the consistency of our new ff$_{\text{O}2}$ relation by comparing the electron temperatures predicted and those directly measured from the spectra.
The bottom right panel of Figure \ref{fig:ff_o2_3d} shows that our ff$_{\text{O}2}$ relation predicts \Te\oii with good precision both for the stacks (red circles) and the single galaxies (blue points), even though single galaxies show a larger scatter from the equality line (in black) as the result of their intrinsic dispersion in the plane which define the ff$_{\text{O}2}$ relation, with a few points whose temperature predictions deviate more than $1000$ K from those observed. 
In addition, we can also compare the temperature prediction of the ff$_{\text{O}2}$ relation with that from the \trel of equation \ref{eq:t2-t3} (applied only to stacks with direct measurement of \Te\oiii), for which the comparison with the direct \Te\oii is shown in the bottom left panel of Figure \ref{fig:ff_o2_3d}.
Our new ff$_{\text{O}2}$ relation clearly reproduces the observed \Te\oii better than the \trel both for our stacks and the \citet{Pilyugin:2010ab} galaxies, as expected given the considerations made in the previous section about how the \trel underestimates the temperature of the low ionization zone when measured from global galaxy spectra. 
For these reasons, in this work we decided to use the new ff$_{\text{O}2}$ relation defined by equation \ref{eq:my_ff}, instead of the \trel, to infer the temperature of the low ionization zone in single, low metallicity galaxies where a direct measurement of \Te\oii was not available (see Sect. \ref{Sec:calib}).

Summarizing, temperatures are derived as follows: when both \oiii$\lambda4363$ and \oii$\lambda7320$ are detected, t$_{2}$ and t$_{3}$ are computed directly from the diagnostic ratios involving these auroral lines; when one of the two lines is missing, we use the relative 
ff relation to infer the flux of that line, compute the diagnostic ratio and derive \Te.
We do not rely in this work on any relation, either empirically derived or based on photoionization models calculations, which links the temperatures of the different ionization zones.

\subsection{Ionic Abundances}
\label{sec:ionic}
\begin{figure}
\centering	
\includegraphics[width=1.\columnwidth, trim = 0.5cm 1cm 0.5cm 0cm]{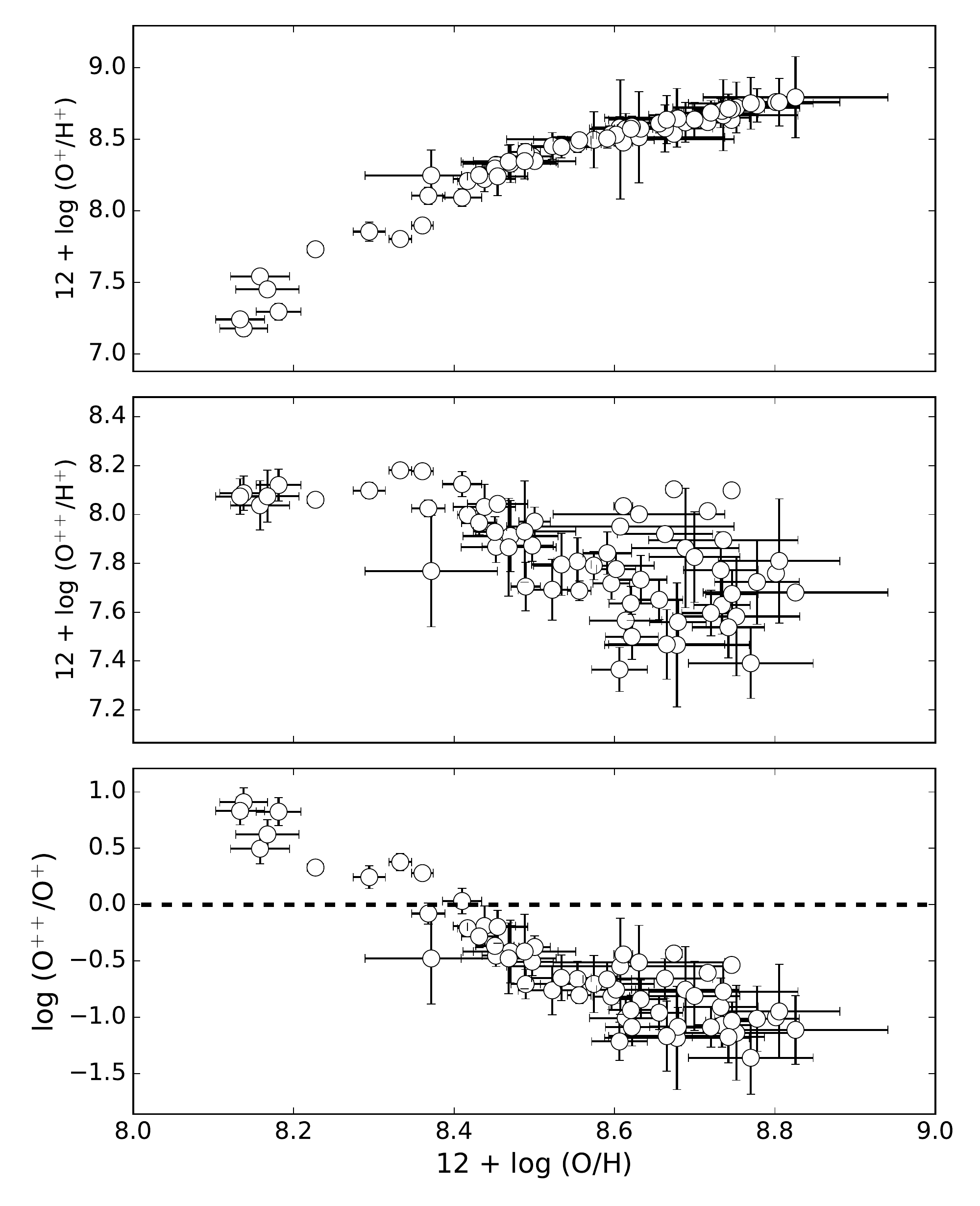}
\caption{The O$^{+}$ abundance (upper panel), the O$^{++}$ abundance (middle panel) and the relative ionic abundance of the two species (bottom panel) are shown as a function of the total oxygen abundance.}
\label{fig:relative_abundances}
\end{figure}

We have calculated the ionic abundances of O$^{+}$ and O$^{++}$ for our stacks with the Pyneb version of the IRAF \emph{nebular.ionic} routine, which determines the abundance of a ionic species given the electron temperature, electron density and the flux ratio of the relative strong emission line with respect to 
H$\beta$.   
We then assume that the total oxygen abundance is the sum of the two species considered, 
\begin{equation}
\frac{\text{O}}{\text{H}} = \frac{\text{O}^{+}}{\text{H}^{+}} + \frac{\text{O}^{++}}{\text{H}^{+}}\ ,
\end{equation}
neglecting the contribution from O$^{3+}$, that can be found in highly ionized gas but it is typically minimal \citep{Andrews:2013ol}.
In calculating the O$^{+}$ abundance we used the electron temperature derived from the \oii diagnostic ratios, while to derive the O$^{++}$ abundance we used the electron temperature derived from the \oiii diagnostic ratios.
For stacks with undetected \oiii$\lambda 4363$ we used the ff$_{\text{O}3}$ relation of equation \ref{eq:ff_rel} to infer \Te \oiii and compute the O$^{++}$ abundance of unseen ionization states.
The systematic offset introduced in abundance determination is small, as we can see by comparing the total oxygen abundance inferred both from \Te \oiii and from the ff$_{\text{O}3}$ relation in stacks with \oiii$\lambda 4363$ detection.
The mean offset in metallicity is $0.028$ dex, smaller than the typical abundance uncertainty.
For comparison, the \trel introduces an average metallicity overestimation of $0.19$ dex, as a direct consequence of the underestimation of the \Te [O{\sc iii}]. 
However, since the relative contribution of the O$^{+}$ state is dominant in almost all the stacks, especially in the high metallicity region, the inferred  O$^{++}$ represent only a small contribution to the total oxygen abundance.
This is shown in Figure \ref{fig:relative_abundances}, where the single ionic abundances for the O$^{++}$ and O$^{+}$ species and their ratio are plotted as a function of the total oxygen abundance: a large part of our stacks above 12 + log(O/H) $=8.5$ presents an O$^{++}$ contribution to total oxygen abundance that does not exceed the $10-20 \%$.
The upper and middle panels of Figure \ref{fig:relative_abundances} show how the O$^{+}$ abundance is seen increasing monotonically in our stacks with the total metallicity, while the O$^{++}$ seems to remain constant or slightly decrease, being in any case affected by a large scatter.
A very similar trend was found also by \cite{Andrews:2013ol} (see Figure $5$ of their paper).
We note that in almost all the stacks with significant contribution of the O$^{++}$ abundance (i.e. $\geq 50\%$) we were able to measure \Te\oiii directly.
Uncertainties on ionic abundances were evaluated following the same 
Monte Carlo simulations used to compute errors on electron temperatures.

\section{Tests on the method} 
\label{Sec:stack_tests}

In this work we stacked spectra of several hundreds of galaxies per bin in order to enhance the signal-to-noise ratio and detect the auroral lines needed for the application of the \Te\ method.
Of course, physical properties like electron temperatures and metallicities inferred from stacked spectra are meaningful only if they are a good representation of the average properties of objects that went into the stack.
In particular, the risk is that few objects could dominate the contribution on auroral line fluxes, thus biasing the estimate of electron temperature, and consequently of metallicity, from stacked spectra.
This work is also based on the assumption that galaxies with similar values for both \oii$\lambda 3727$/H$\beta$ and \oiii$\lambda 5007$/H$\beta$ have similar metallicities.

\begin{figure*}
\centering
\includegraphics[width=2.\columnwidth, trim = 0.5cm .4cm .5cm .4cm]{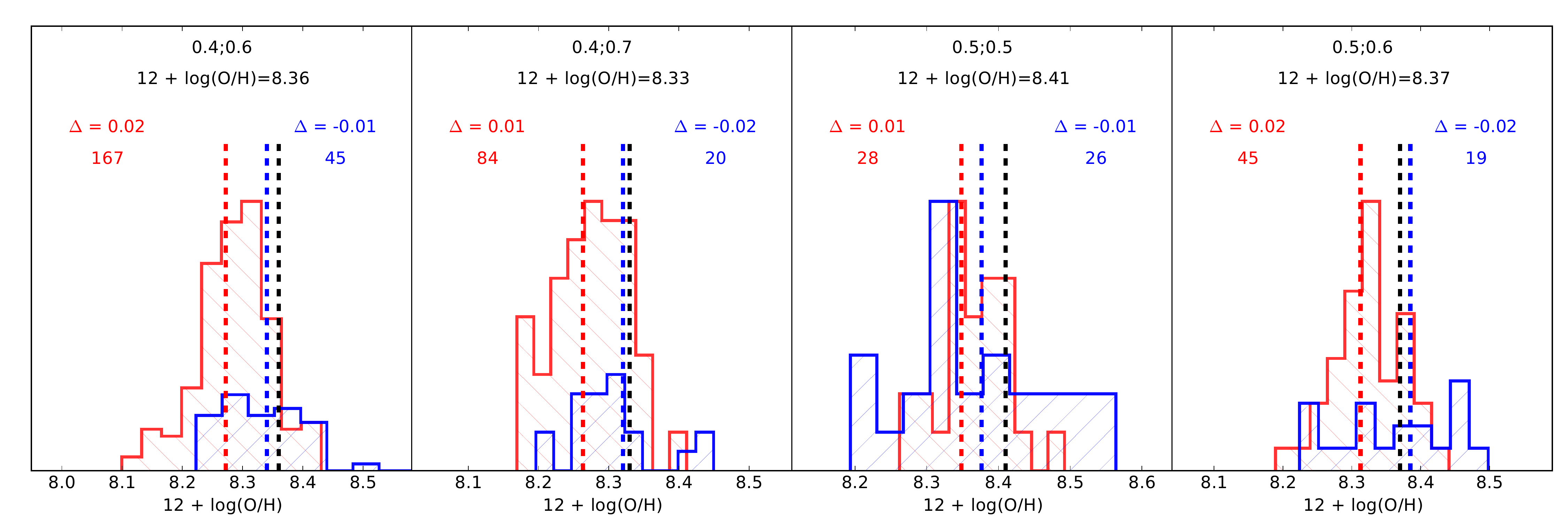}
\caption{Histograms of \Te\ method metallicities for the subsample of galaxies selected from \citet{Pilyugin:2010ab} with detected \oiii$\lambda 4363$ and \oii$\lambda 7320,7330$ auroral lines (blue sample) and for galaxies with \oiii$\lambda 4363$ detected at $>10\sigma$ from the MPA/JHU catalog (red sample) for the $0.4;0.6$, $0.4;0.7$, $0.5;0.5$ and $0.5;0.6$ bin.
The dashed lines indicate the metallicity inferred from the composite spectra obtained stacking the relative sample of galaxies.
In every panel are also reported, for both subsamples, the difference between the average metallicity of the distribution and the value inferred from the associated stacked spectrum ($\Delta$) and the number of objects per stack. The metallicity of the global stack, i.e. the stack obtained from the full sample of galaxies that fall in that bin, is written in the upper part of each panel and indicated by the dashed black line.} 
\label{fig:4363_hist}
\end{figure*}

In order to test our hypothesis and the stacking procedure, we took the sample of galaxies from \cite{Pilyugin:2010ab} with detected \oiii$\lambda 4363$ and \oii$\lambda 7320,7330$, and stacked these spectra in bins of \oii/H$\beta$ and \oiii/H$\beta$ according to our pipeline.
For this analysis we considered only those bins with at least $15$ objects, which namely are $0.4;0.6$, $0.4;0.7$, $0.5;0.5$ and $0.5;0.6$, as reported at the top of Figure \ref{fig:4363_hist}.
Furthermore, we searched for those galaxies in our original sample (described in Section 2), with detection of \oiii$\lambda 4363$ in the MPA/JHU catalog, falling into the same bins. In particular, we selected only galaxies with \oiii $\lambda 4363$ detected at $\geq 10 \sigma$.

Temperatures and metallicities were computed for each single galaxy in both samples. The sample of \citet{Pilyugin:2010ab} galaxies has both the oxygen auroral lines detected, thus we were able to directly infer the temperatures and abundances of both oxygen ionic species.
Since only the \oiii $\lambda 4363$ auroral line is instead available for objects selected from the MPA/JHU catalog, we used the ff$_{\text{O}2}$ relation of equation \ref{eq:my_ff} to determine the flux of the \oii$\lambda 7320,7330$ auroral doublet and compute \Te\oii and the O$^{+}$ abundance.
Then, we generated, for each bin, separate composite spectra for both samples and we measured T$_{e}$ and metallicities with the \Te\ method from the stacked spectra.

Figure \ref{fig:4363_hist} shows the histograms of metallicity distribution of individual galaxies in each bin for the \cite{Pilyugin:2010ab} sample (from now on: the blue sample) and the sample selected from the MPA/JHU catalog (from now on: the red sample). 
We note the quite small range of metallicities spanned by single galaxies in each bin, with typical dispersions of $0.1$ dex, consistently with the width of our binning grid. 
Even though we can not perform the same test for higher metallicity stacks due to the lackness of auroral line detection in single galaxies, this corroborates the assumption that galaxies belonging to a given bin of fixed \oiii$\lambda 5007$/H$\beta$ and \oii$\lambda 3727$/H$\beta$ have similar metallicities and that we are thus stacking objects with similar properties in terms of oxygen abundance.
The dashed lines in Figure \ref{fig:4363_hist} indicate instead the metallicity inferred from the associated stacked spectrum for both samples. 
The difference between the average metallicity of single galaxies in a given bin and the one inferred from the stacked spectrum is reported as $\Delta$; the number of objects per bin is also written. 
We note that abundances estimated from stacks are well matched to the average of the metallicity distributions in every bin, with offsets being at most $0.02$ dex for both samples.
However, both the red and the blue sample could not be fully representative of the galaxy population inside each bin, which consist also of a large number of galaxies with no detection of auroral lines.
Therefore, we compare the metallicity inferred from the stacked spectra of both 
sub-samples with the one derived from the global composite spectrum, i.e. the spectrum obtained stacking all the galaxies included in that bin according to the procedure described in Section \ref{Sec:method}. These values are reported at the top of each box of Figure \ref{fig:4363_hist} and indicated by the black dashed lines. 
We find good agreement between the global stack metallicity and the one inferred from the stacked spectra of the two different sub-samples, with typical offsets on average of $0.04$ dex, even though we note a systematic metallicity underestimation when considering the two sub-samples with respect to the global one.
This is probably due to the fact that, when creating the stacked spectra for the different sub samples, we are averaging upon the most metal poor galaxies in the bin, which in fact have the auroral lines detected. 
This could bias the sub-sample stacks toward lower metallicities, but this effect is smaller both than our bin size and than the average uncertainty associated to abundances measurements in our stacks.
We therefore conclude that different sub-sampling criteria inside the same bin does not dramatically affect the metallicity estimation from composite spectra and therefore that stacked spectra are effectively representative of the average properties, in terms of oxygen abundance, of the objects from which they are generated.

\section{Calibrations of strong-line metallicity indicators}
\label{Sec:calib}

In order to extend the metallicity range covered by our calibrations, we add to our stacks a sample of single galaxies with robust detection of \oiii$\lambda 4363$. We selected galaxies from our original SDSS DR 7 sample with \oiii$\lambda 4363$ detection at $> 10 \sigma$, and we re-computed the oxygen abundance for these galaxies according to the procedure described in the previous Section.
In particular, we derive \Te\oiii directly exploiting the \oiii$\lambda 4363$ value reported on the MPA/JHU catalog and used the ff$_{\text{O}2}$ relation of equation \ref{eq:my_ff} to infer \Te\oii and the O$^{+}$ ionic abundance. 
Even though a part of these galaxies, although not all of them, are already included into our stacking grid, we are able in this way to directly account for some of the most metal poor galaxies of our sample, without averaging them into the stacking bins; thus, we can better constrain the low metallicity region of our calibrations. 

\begin{figure*}
\centering
\includegraphics[width=2\columnwidth, trim = 0.5cm 0.cm 0.2cm 0.2cm]{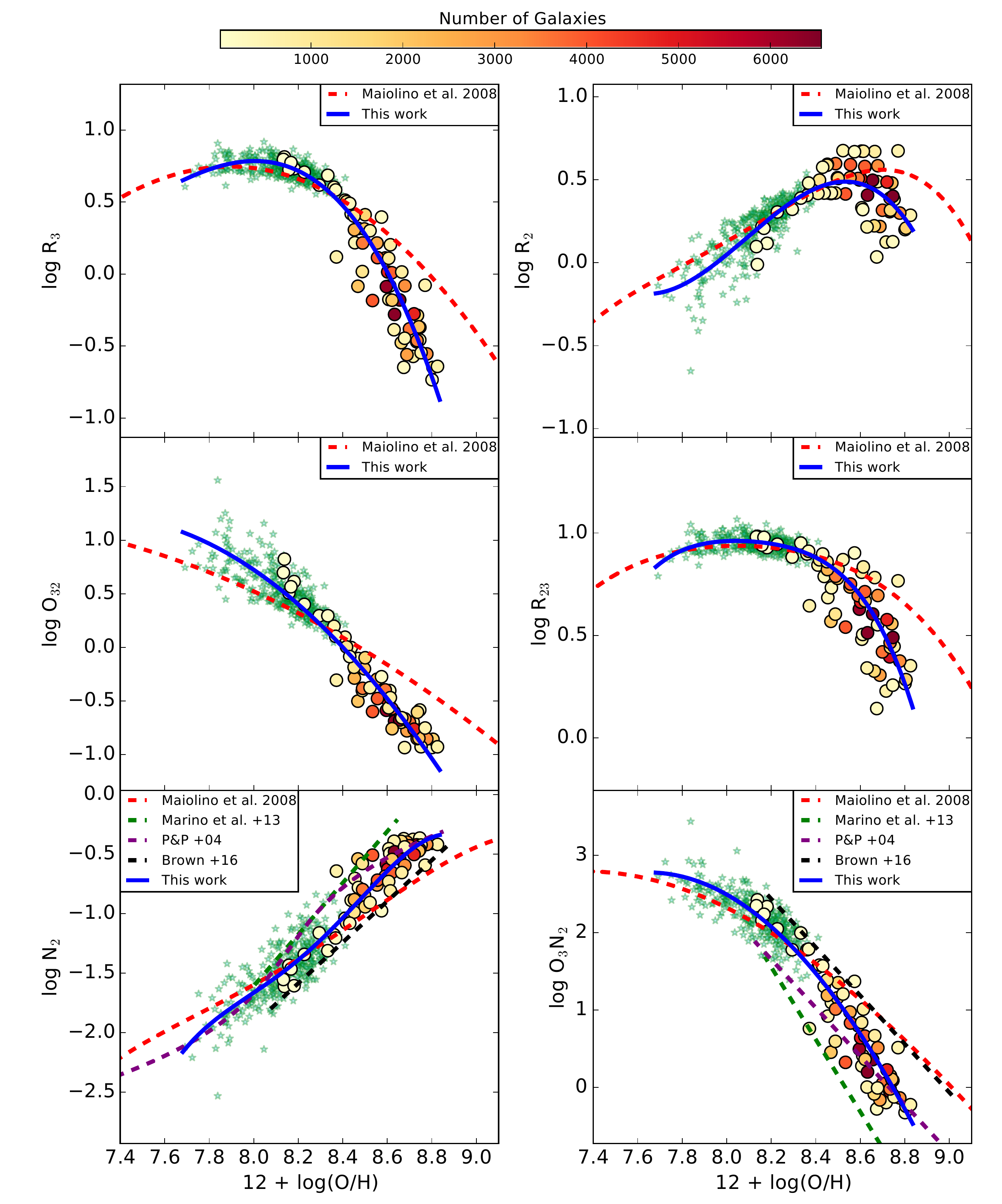}
\caption{Strong Line diagnostics as a function of oxygen abundance for our full sample: small green stars represent the sample of single SDSS galaxies with \oiii$\lambda 4363$ detected at S/N $>10$, while circles are the stacks color coded by the number of galaxies in each bin. Our best fit polynomial functions are shown as solid blue curves, while the dashed red line represents the \citet{Maiolino:2008cv} semi-empirical calibrations. 
In the N$_{2}$ and O$_{3}$N$_{2}$ diagrams also the \citet{Pettini:2004fk}(dashed purple curve), \citet{Marino:2013ty}(dashed green curve) and \citet{Brown:2016aa} for $\Delta$(SSFR)=0 (dashed black curve) 
calibrations are shown.
A publicly available routine to apply these calibrations can be found at \url{http://www.arcetri.astro.it/metallicity/}. 
}
\label{fig:calibrazioni}
\end{figure*}

In Figure \ref{fig:calibrazioni} we plot the relations between some of the most widely used strong-line metallicity indicators and gas-phase oxygen abundance for our full sample.
In particular, we re-calibrate R$_{2}($\oii$\lambda3727$/H$\beta$), R$_{3}$(\oiii$\lambda5007$/H$\beta$), R$_{23}$((\oii$\lambda 3727$+\oiii$\lambda 4959,5007$)/H$\beta$), O$_{32}$ (\oiii$\lambda 5007$/\oii$\lambda 3727$), N$_{2}$(\nii$\lambda 6584$/H$\alpha$) and O$_{3}$N$_{2}$ ((\oiii$\lambda 5007$/H$\beta$)/(\nii$\lambda 6584$/H$\alpha$)).
Green small stars represent single galaxies, 
while circles represent our stacked spectra, color coded by the number of objects that went into each stack.
To derive our new calibrations, we performed a polynomial fitting
whose general functional form is  
\begin{equation}
\label{eq:calib}
\text{log R} = \sum_{N} c_{n} x^{n}
\end{equation}
where R is a given diagnostic and x is the oxygen abundance normalized to the solar value ($12 +\text{log(O/H)}_{\odot}= 8.69$, \citealt{Allende-Prieto:2001aa}). 
Since the indicators based on the ratio between oxygen forbidden lines over hydrogen recombination lines exhibit the well-known double branch behavior, a high order polynomial fitting is required.
Assuming that the uncertainty on the auroral line flux, which represents the main contribution to the error in the \Te\ abundances determination in our stacks, decreases as the square root of the number of galaxies, in our fitting procedure  we assigned a weight equal to this value to each point representing a stack; points associated to single galaxies have been weighted as they were stacks of only one object.
In this way we also avoid our fit to be dominated by the low metallicity single galaxies which are far more numerous than the stacks.
In Figure \ref{fig:calibrazioni} our new calibrations are shown with the blue curve and in Table  \ref{tab:coeff_calib} the best fit coefficients and the RMS of the residuals of the fit are reported for each of them. 

We then applied each calibration to our total sample of single galaxies and stacks and computed the differences between \Te\ method metallicity and metallicity predicted by the calibration, in order to give an estimate of the dispersion along the log(O/H) direction, which is reported as $\sigma$ in Table \ref{tab:coeff_calib}. 
For double branched diagnostics (i.e. R$_{3}$, R$_{2}$ and R$_{23}$) this estimate is provided only considering the metallicity range where they show monotonic dependence on log(O/H), which is reported in the \textit{Range} column of Table \ref{tab:coeff_calib}.
This column represents indeed the range of applicability for a given diagnostics when used as single metallicity indicator.
We note that $\sigma$ should not be directly interpreted as the uncertainty to associate to metallicity determination with our calibrations, since uncertainties in emission line ratios could introduce comparable errors.

Since our calibrations are build from a non homogeneous combination of single galaxies and stacks, dispersion in our diagrams is due to different contributions. 
In the range covered by single SDSS galaxies, it is the consequence of the intrinsic spread in a given strong line ratio at fixed metallicity and of the uncertainty on the auroral line fluxes measurement.
For the high metallicity region covered by our stacks, since we are averaging on a large number of objects, the scatter due to the intrinsic dispersion should be in principle reduced.
However, we must consider the effects associated with the particular choice of our stacking grid.
Every stack has, by definition, a defined value of [O{\sc ii}]/H$\beta$ and \oiii/H$\beta$; therefore, in the R$_{2}$ and  R$_{3}$ calibration diagrams the residual dispersion reflects the segregation in a given diagnostic when the other is fixed. This means that for any given value of one line ratio, different metallicities can be found varying the other one.
This is particularly clear in the R$_{2}$ calibration, where different sequences for different [O{\sc iii}]/H$\beta$ values appears at metallicities above $8.2$.
Therefore, this diagnostic shows a clear dependence on oxygen abundance only in the low metallicity regime, revealing how most of SDSS galaxies are falling in the transition zone between the two branches of this indicator. 
Thus, for the majority of our stacks the metallicity dependence is driven by [O{\sc iii}]/H$\beta$, and indeed for this diagnostic the segregation in sequences of [O{\sc ii}]/H$\beta$ is much less prominent.

\begin{table*}
\caption{Best fit coefficients and RMS of the residuals for calibrations of metallicity diagnostics given by equation \ref{eq:calib}. The $\sigma$ parameter is an estimate of the dispersion along the log(O/H) direction in the interval of applicability given in the \textit{Range} column.}
\label{tab:coeff_calib}
\medskip
\centering
\begin{tabular}{lcccccccc}
\toprule
Diagnostic  	& c$_{0}$   & c$_{1}$      & c$_{2}$      & c$_{3}$      & c$_{4}$    & RMS & $\sigma$  & Range  \\
\midrule
 R$_{2}$		&	0.418	&		-0.961	&	-3.505	    &   -1.949	   &  		    &    0.11	  &		0.26    &   7.6 < 12+log(O/H) < 8.3   	\\
 R$_{3}$		&	-0.277	&		-3.549  &	-3.593		&	-0.981 	   &   		    &	 0.09	  &	    0.07	&   8.3 < 12+log(O/H) < 8.85   \\
 O$_{32}$		&	-0.691	&	-2.944		&	-1.308		&		   	   & 			&	 0.15	  &	    0.14	&      7.6 < 12+log(O/H) < 8.85 \\
 R$_{23}$		&	0.527	&	-1.569	    &	-1.652	    &	-0.421	   & 			&	 0.06	  &		0.12	&   8.4 < 12+log(O/H) < 8.85   \\
 N$_{2}$		&	-0.489	&	1.513		&	-2.554		&	-5.293	   &	-2.867	&	 0.16     &		0.10	&      7.6 < 12+log(O/H) < 8.85 \\
 O$_{3}$N$_{2}$	&	0.281	&	-4.765		&	-2.268		&		       & 	        &	 0.21	  &		0.09    &      7.6 < 12+log(O/H) < 8.85  \\
\bottomrule
\end{tabular}
\end{table*}

For other indicators, the dispersion mainly reflects the scatter for a given diagnostic line ratio inside each [O{\sc ii}]/H$\beta$-[O{\sc iii}]/H$\beta$ bin. For each diagnostic the distribution of the corresponding line ratio inside our bins is generally strongly peaked, even though we are affected by different dispersions when considering different positions on our stacking grid. This means that a given line ratio, as measured from the stacked spectra, can be respectively more or less representative of the distribution of galaxies inside a given bin for different positions on the diagram.
However, for every diagnostic ratio here considered, the typical dispersion of its distribution inside a given bin is of the order of $0.1$ dex (or less), thus being consistent with the choice of our bin size.

In Figure \ref{fig:calibrazioni} we compare our new calibrations with those from \cite{Maiolino:2008cv}.
They obtained semi empirical calibrations combining direct abundance determination for galaxies from the \cite{Nagao:2006gd} sample with metallicity estimation from theoretical models by \cite{Kewley:2002aa}. 
The two calibrations agree well, as expected, for most of the indicators at low metallicities, the main discrepancies arising in the high metallicity regime where \Te\ method metallicities of our stacks result lower than those predicted by photoionization models. This introduce a clear deviation in the slope in all our calibrations, that change significantly their steepness after $12 + \text{log(O/H)} \sim 8.2$.
In fact, we note that the highest metallicities inferred from our composite spectra are only slightly higher ($\sim 0.1$ dex) than the solar value.

For the O$_{3}$N$_{2}$ and N$_{2}$ indicators we can compare our calibrations also with empirical ones from \citet{Pettini:2004fk} and \citet{Marino:2013ty}, who used single \Hii regions and not integrated galaxy spectra to calibrate these line ratios against metallicity.  Our calibrations have comparable slopes to those of \cite{Marino:2013ty}, but they present a systematic offset towards higher metallicities. This is probably due to the fact that calibrations entirely based on \Hii regions like \cite{Marino:2013ty} are biased towards high excitation conditions and low metallicities.
Our N$_{2}$ calibration is in good agreement with \citet{Pettini:2004fk}  at low metallicities but diverge, in the direction of predicting higher abundances, in the middle region.  At metallicities close to solar this diagnostic begin to saturate, as expected from the fact that nitrogen becomes the dominant coolant of the ISM: the two calibrations then become comparable again.
The O$_{3}$N$_{2}$ calibration instead presents a different slope than the \citet{Pettini:2004fk} since the slope of their calibration is determined by the use of photoionization models at high metallicities due to the lack in their sample of \Hii regions with direct abundances in that region of the diagram.
We note that our calibrations are better constrained to be used for integrated galaxy spectra, since single \Hii regions upon which most of the empirical calibrations are based on do not properly and fully cover the parameter space where many galaxies lie.

In Figure \ref{fig:calibrazioni} we also compare our calibrations for the N$_{2}$ and O$_{3}$N$_{2}$ indicators with those derived by \citet{Brown:2016aa}, who derived oxygen abundances with the \Te\ method from stacked spectra in bins of stellar mass and $\Delta$(SSFR), i.e. the deviation of the specific star formation rate from the star forming main sequence (SFMS) \citep{Noeske:2007aa}. Since they include $\Delta$(SSFR) as a second parameter in their calibrations, we decide to plot here (in black) only the curves representative of SFMS galaxies, i.e those obtained assuming $\Delta$(SSFR)=0. In fact, our galaxy sample is distributed around their SFMS representation (see equation $6$ of \citealt{Brown:2016aa}), with a small median 
offset of $0.009$ dex.
Their calibrations show an offset of $\sim 0.1$ dex towards higher metallicities for both indicators with respect to ours. We note that our calibrations are more consistent with \cite{Brown:2016aa} calibrations when considering their curves for $\Delta$(SSFR)$=-0.75$.

\begin{figure}
\centering
\includegraphics[width=1\columnwidth, trim = .2cm 0.5cm .2cm 0.3cm]
{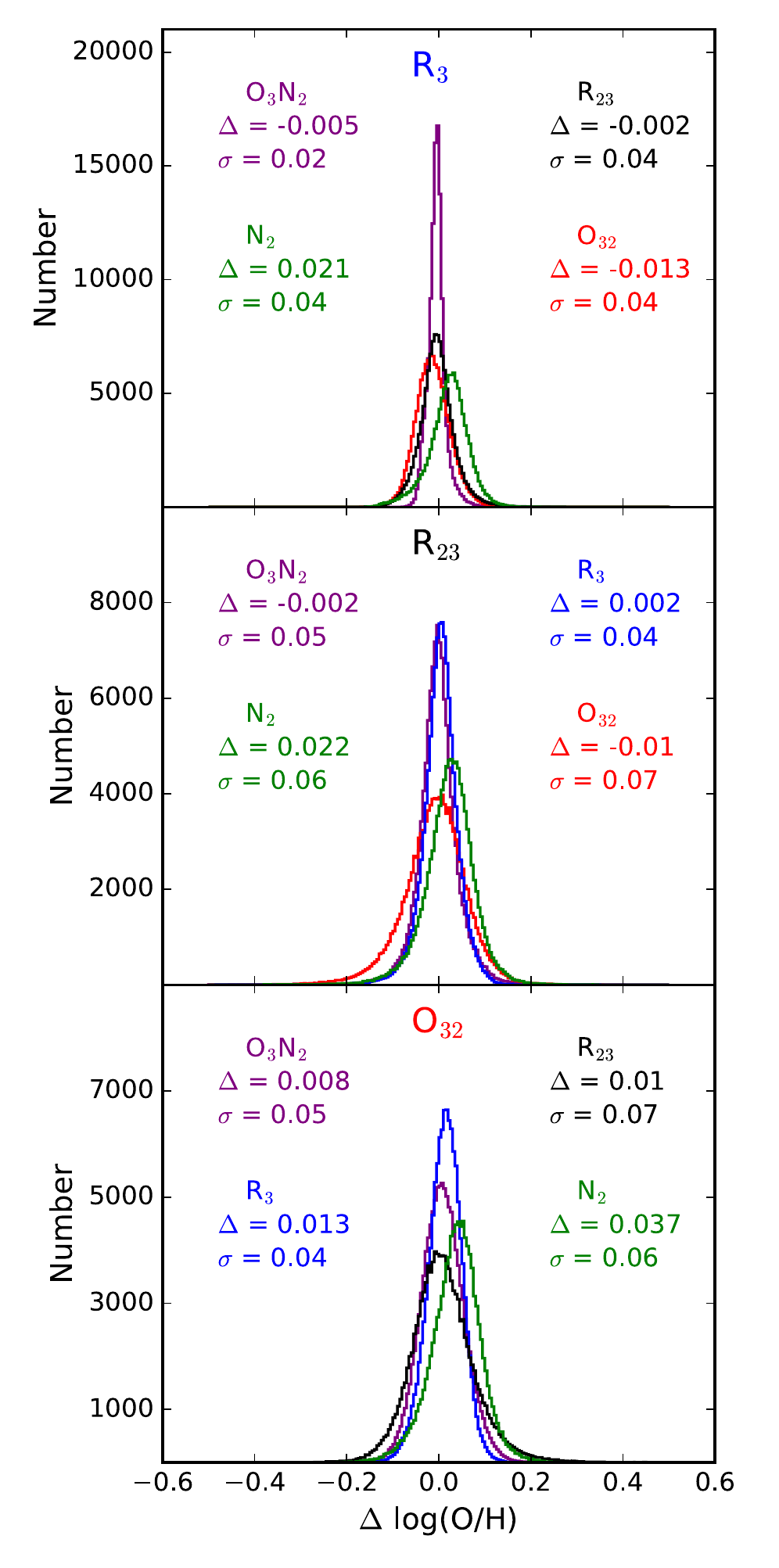}
\caption{\textit{Upper Panel} : Histograms of the difference between metallicities of the SDSS galaxies derived with the R$_{3}$ calibration and through the other diagnostics. Each diagnostic is identified by a different color:
blue for  R$_{3}$, red for  O$_{32}$, black for  R$_{23}$, purple for  O$_{3}$N$_{2}$ and green for  N$_{2}$. The average offset and sigma of the $\Delta$(O/H) distributions is written for every diagnostic with the associated color.
\textit{Middle Panel} : Same as Upper Panel, with R$_{23}$ as reference diagnostic. \textit{Bottom Panel} : Same as Upper Panel, with O$_{32}$ as reference diagnostic.}
\label{fig:histo_cal1}
\end{figure}
\begin{figure}
\centering
\includegraphics[width=1\columnwidth, trim = .3cm 0.5cm .3cm .7cm]
{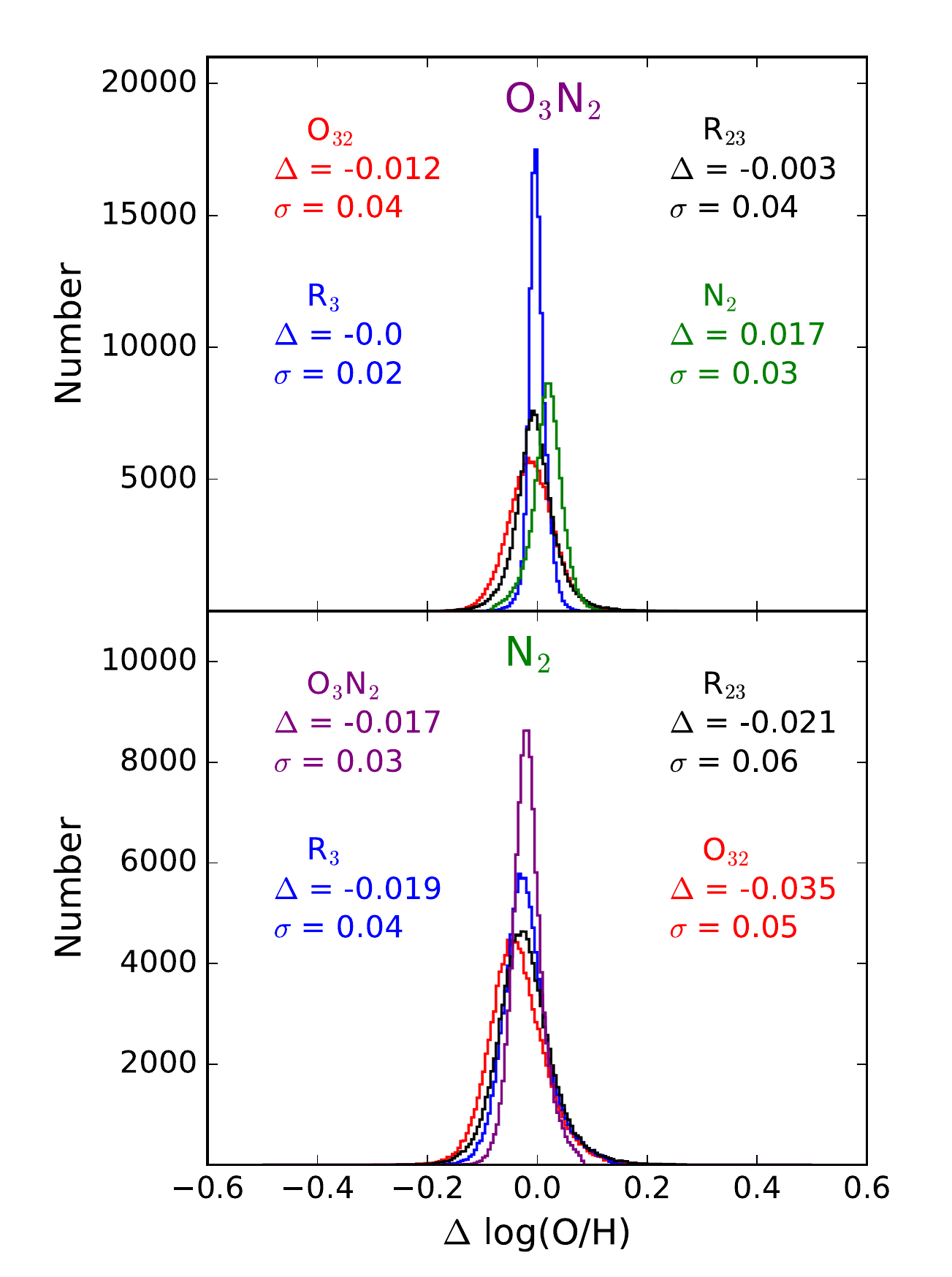}
\caption{\textit{Upper Panel} : Same as Upper Panel of Figure \ref{fig:histo_cal1}, with O$_{3}$N$_{2}$ as reference diagnostic.
\textit{Bottom Panel} : Same as Upper Panel of Figure \ref{fig:histo_cal1}, with N$_{2}$ as reference diagnostic.}
\label{fig:histo_cal2}
\end{figure}

In order to test the self-consistency of our calibrations, we applied them individually to our original sample of SDSS galaxies.
Diagnostics behaving monotonically (i.e. N$_{2},$O$_{3}$N$_{2}$ and O$_{32}$) can be compared over the full metallicity range spanned by our sample, to which we can straightly apply the calibration, while for those double valued we restricted our analysis to the interval given in the \textit{Range} column of Table \ref{tab:coeff_calib}.
Given that our R$_{2}$ calibration mostly cover the transition zone for such indicator in our SDSS galaxy sample (i.e. for $12 + \text{log(O/H)} > 8.2$), we decide not to include it in this analysis.
In each panel of Figure \ref{fig:histo_cal1} and \ref{fig:histo_cal2} we show the histograms of the differences in metallicity estimation between a given diagnostic and all the others.
Every strong-line indicator is identified by a different colour, and  
in each panel the name of the reference indicator is written in the upper region. The $\Delta$log(O/H) is then evaluated as the difference between the metallicity probed by the reference indicator and the metallicities estimated with the other four; in each panel the average offset of the $\Delta$log(O/H) distribution and the associated $\sigma$ are reported.

Inspection of the different panels of Figures \ref{fig:histo_cal1} and \ref{fig:histo_cal2} reveals that metallicities probed by different indicators are in good agreement among each other, with average offsets below $0.04$ dex and typical dispersions below $0.1$ dex.
The little systematic metallicity overestimate reported for the O$_{32}$ indicator and underestimate for the N$_{2}$ indicator with respect to the others can be accounted for as a product of the fitting procedure. 
In this sense, the use of higher order polynomials allow to straighten the consistency of all our calibrations, since it minimizes the mutual disagreement between metallicity determined with different indicators.
Thus, Figures \ref{fig:histo_cal1} and \ref{fig:histo_cal2} show that consistent metallicities are obtained in individual galaxies when using different calibrations, allowing to compare for example  abundances obtained from diagnostics located in different spectral regions.

From the above considerations, we can say that our calibrations represent a self consistent set totally based on the \Te\ metallicity scale. 
This is important since at the present time there is not an established absolute abundance 
scale for galaxies. 
Most of the calibrations found in literature either rely on the assumptions of photoionization models or are based on individual \Hii regions metallicities. In the first case the metallicity scale defined by models is inconsistent with the \Te\ scale. In the second case the emission lines properties of \Hii regions differs from those of integrated galaxy spectra and higher excitation conditions affect \Te\ abundances estimation towards lower values.
Until the number of high metallicity galaxy with detected auroral lines will increase, allowing to build fully \Te\ method calibrations based on samples of individual objects, our stacking technique represents a valuable approach to define \Te\ based calibrations.
However, it should be stressed that metallicity estimates obtained from these calibrations are always affected by the particular choice of the stacking procedure.
For example, the \citet{Brown:2016aa} calibrations from stacked spectra in bins of stellar mass and $\Delta$(SSFR) rely on a particular parametrization of the local star forming main sequence and their metallicity predictions could differ from ours despite the fact that abundances are evaluated with the \Te\ method in both cases. 

In this work we chose to re-calibrate the strong-line diagnostics relying only on the values assumed by galaxies on particular emission line ratios, thus assuming only the validity of the strong-line-methods to infer metallicity from spectra of star forming galaxies.
In this way our calibrations could be in principle applied to a great variety of cases, in particular to large IFU galaxy surveys that allow spatially resolved abundance studies (e.g CALIFA, \citealt{Sanchez:2012aa}, MaNGA, \citealt{Bundy:2015aa} or SAMI, \citealt{Croom:2012aa}). 
In such cases for example, the use of our calibrations allows to relax the assumption that scaling relations well assessed on global scales (e.g. the M-Z and the M-Z-SFR relations) still hold on smaller, local ones.

\section{Summary}
\label{sec:summary}

We provided new and totally empirical calibrations for some of the most widely used strong-line diagnostics for the determination of oxygen abundance in star forming galaxies.
These relations have been derived combining a sample of single low metallicity galaxies together with the stacking of more than $110\,000$ galaxies of the SDSS in bins of $0.1$ dex in the log \oii$\lambda3737$/H$\beta$ - log \oiii$\lambda5007$/H$\beta$ diagram, just assuming that galaxies with such similar strong line ratios also show similar metallicity (i.e. assuming the validity of the so called Strong Line Method).
The increase in signal-to-noise ratio provided by the stacking procedure allowed us to detect and measure both the \oiii $\lambda 4363$ and \oii $\lambda 7320,7330$ auroral lines necessary to compute electron temperatures of the different ionization zones and apply the \Te\ method for measuring metallicity on the full range of metal abundances spanned by galaxies in the SDSS survey. Here are summarized our main results :
\begin{itemize}

\item{We found evidence for [Fe {\sc ii}] contamination of the \oiii$\lambda 4363$ auroral line in high metallicity stacks (Figure \ref{fig:4363_flag}) . 
This is one of the crucial lines for the application of the \Te\ method for abundances estimation, thus we recommend care in using this line as electron temperature diagnostic when detected in high metallicity (12 + log(O/H) $\gtrsim 8.3$) galaxy spectra. 
}
\item{We analysed the relations between electron temperatures of different ionization zones, finding that our stacks do not follow the  established \trel for \Hii regions (Figure \ref{fig:t2_t3_diagrams}) . They show instead better agreement with a relation that correlates auroral and nebular line fluxes (ff$_{\text{O}3}$ relation).  
Exploiting the direct detection of the \oii$\lambda 7320,7330$ auroral doublet in all of our composite spectra, we provided a new relation (the ff$_{\text{O}2}$ relation) for the determination of the flux of this auroral line (Figure \ref{fig:ff_o2_3d}).}
\item{
We then analysed the relations between some of the most common strong line diagnostics and oxygen abundance in order to obtain a reliable calibration (Figure \ref{fig:calibrazioni}). Our global sample allowed us to construct a set of calibrations, spanning more than $1$ dex in metallicity, which are based on the uniform application of the \Te\ method for oxygen abundances estimation on global galaxy spectra.
All our calibrations are therefore defined, over their whole range, on a consistent absolute \Te\ metallicity scale for local star forming galaxies. The scatter around the best fitting calibration varies between $0.05$ and $0.15$ for different indicators.}

\item{
Comparing our new calibrations with different ones from literature reveals 
how our calibrations deviate significantly 
both from empirical ones based on \Hii regions and from theoretical ones based on photoionization models, especially at high metallicities. In fact, we find that our most metal rich stacks have oxygen abundance significantly lower than those predicted by models, and at most $0.14$ dex higher than the solar one (i.e. $\sim 1.4\ \text{Z}_{\odot}$). On the other hand, classical empirical calibrations obtained from \Hii regions samples generally show lower metallicities for fixed line ratios,
probably due to the fact that the single \Hii regions used for those calibrations are somehow biased towards high excitation conditions in order to ensure auroral line detection at high metallicity.}

\item{We applied our calibrations to the original sample of SDSS galaxies.
Metallicity estimates from different calibrations result in good agreement between each other, with typical average offsets lower than 
$0.04$ dex and dispersions of the order of $0.05$ dex.
We do not find any systematic effect of metallicity overestimate or underestimate between the different diagnostics. Thus, our calibrations represent a self consistent set that could be used in a variety of different cases depending on the availability of emission lines.}
\end{itemize}

\bibliographystyle{mnras}

\begin{thebibliography}{}
\makeatletter
\relax
\def\mn@urlcharsother{\let\do\@makeother \do\$\do\&\do\#\do\^\do\_\do\%\do\~}
\def\mn@doi{\begingroup\mn@urlcharsother \@ifnextchar [ {\mn@doi@}
  {\mn@doi@[]}}
\def\mn@doi@[#1]#2{\def\@tempa{#1}\ifx\@tempa\@empty \href
  {http://dx.doi.org/#2} {doi:#2}\else \href {http://dx.doi.org/#2} {#1}\fi
  \endgroup}
\def\mn@eprint#1#2{\mn@eprint@#1:#2::\@nil}
\def\mn@eprint@arXiv#1{\href {http://arxiv.org/abs/#1} {{\tt arXiv:#1}}}
\def\mn@eprint@dblp#1{\href {http://dblp.uni-trier.de/rec/bibtex/#1.xml}
  {dblp:#1}}
\def\mn@eprint@#1:#2:#3:#4\@nil{\def\@tempa {#1}\def\@tempb {#2}\def\@tempc
  {#3}\ifx \@tempc \@empty \let \@tempc \@tempb \let \@tempb \@tempa \fi \ifx
  \@tempb \@empty \def\@tempb {arXiv}\fi \@ifundefined
  {mn@eprint@\@tempb}{\@tempb:\@tempc}{\expandafter \expandafter \csname
  mn@eprint@\@tempb\endcsname \expandafter{\@tempc}}}

\bibitem[\protect\citeauthoryear{{Abazajian} et~al.,}{{Abazajian}
  et~al.}{2009}]{Abazajian:2009aa}
{Abazajian} K.~N.,  et~al., 2009, \mn@doi [\apjs]
  {10.1088/0067-0049/182/2/543}, \href
  {http://adsabs.harvard.edu/abs/2009ApJS..182..543A} {182, 543}

\bibitem[\protect\citeauthoryear{{Aggarwal} \& {Keenan}}{{Aggarwal} \&
  {Keenan}}{1999}]{Aggarwal:1999aa}
{Aggarwal} K.~M.,  {Keenan} F.~P.,  1999, \mn@doi [\apjs] {10.1086/313232},
  \href {http://adsabs.harvard.edu/abs/1999ApJS..123..311A} {123, 311}

\bibitem[\protect\citeauthoryear{{Allende Prieto}, {Lambert}  \&
  {Asplund}}{{Allende Prieto} et~al.}{2001}]{Allende-Prieto:2001aa}
{Allende Prieto} C.,  {Lambert} D.~L.,   {Asplund} M.,  2001, \mn@doi [\apjl]
  {10.1086/322874}, \href {http://adsabs.harvard.edu/abs/2001ApJ...556L..63A}
  {556, L63}

\bibitem[\protect\citeauthoryear{{Alloin}, {Collin-Souffrin}, {Joly}  \&
  {Vigroux}}{{Alloin} et~al.}{1979}]{Alloin:1979lq}
{Alloin} D.,  {Collin-Souffrin} S.,  {Joly} M.,   {Vigroux} L.,  1979, \aap,
  \href {http://adsabs.harvard.edu/abs/1979A%26A....78..200A} {78, 200}

\bibitem[\protect\citeauthoryear{{Andrews} \& {Martini}}{{Andrews} \&
  {Martini}}{2013}]{Andrews:2013ol}
{Andrews} B.~H.,  {Martini} P.,  2013, \mn@doi [\apj]
  {10.1088/0004-637X/765/2/140}, \href
  {http://adsabs.harvard.edu/abs/2013ApJ...765..140A} {765, 140}

\bibitem[\protect\citeauthoryear{{Berg}, {Croxall}, {Skillman}, {Pogge},
  {Moustakas}  \& {Groh-Johnson}}{{Berg} et~al.}{2015}]{Berg:2015aa}
{Berg} D.~A.,  {Croxall} K.~V.,  {Skillman} E.~D.,  {Pogge} R.~W.,  {Moustakas}
  J.,   {Groh-Johnson} M.,  2015, preprint, \href
  {http://adsabs.harvard.edu/abs/2015arXiv150102270B} {} (\mn@eprint {arXiv}
  {1501.02270})

\bibitem[\protect\citeauthoryear{{Binette}, {Matadamas}, {H{\"a}gele},
  {Nicholls}, {Magris C.}, {Pe{\~n}a-Guerrero}, {Morisset}  \&
  {Rodr{\'{\i}}guez-Gonz{\'a}lez}}{{Binette} et~al.}{2012}]{Binette:2012aa}
{Binette} L.,  {Matadamas} R.,  {H{\"a}gele} G.~F.,  {Nicholls} D.~C.,  {Magris
  C.} G.,  {Pe{\~n}a-Guerrero} M.~{\'A}.,  {Morisset} C.,
  {Rodr{\'{\i}}guez-Gonz{\'a}lez} A.,  2012, \mn@doi [\aap]
  {10.1051/0004-6361/201219515}, \href
  {http://adsabs.harvard.edu/abs/2012A%26A...547A..29B} {547, A29}

\bibitem[\protect\citeauthoryear{{Blanc}, {Kewley}, {Vogt}  \&
  {Dopita}}{{Blanc} et~al.}{2015}]{Blanc:2015aa}
{Blanc} G.~A.,  {Kewley} L.,  {Vogt} F.~P.~A.,   {Dopita} M.~A.,  2015, \mn@doi
  [\apj] {10.1088/0004-637X/798/2/99}, \href
  {http://adsabs.harvard.edu/abs/2015ApJ...798...99B} {798, 99}

\bibitem[\protect\citeauthoryear{{Bresolin}, {Schaerer}, {Gonz{\'a}lez Delgado}
   \& {Stasi{\'n}ska}}{{Bresolin} et~al.}{2005}]{Bresolin:2005aa}
{Bresolin} F.,  {Schaerer} D.,  {Gonz{\'a}lez Delgado} R.~M.,   {Stasi{\'n}ska}
  G.,  2005, \mn@doi [\aap] {10.1051/0004-6361:20053369}, \href
  {http://adsabs.harvard.edu/abs/2005A%26A...441..981B} {441, 981}

\bibitem[\protect\citeauthoryear{{Bresolin}, {Kudritzki}, {Urbaneja}, {Gieren},
  {Ho}  \& {Pietrzynski}}{{Bresolin} et~al.}{2016}]{Bresolin:2016aa}
{Bresolin} F.,  {Kudritzki} R.-P.,  {Urbaneja} M.~A.,  {Gieren} W.,  {Ho} I.,
  {Pietrzynski} G.,  2016, preprint, \href
  {http://adsabs.harvard.edu/abs/2016arXiv160706840B} {} (\mn@eprint {arXiv}
  {1607.06840})

\bibitem[\protect\citeauthoryear{{Brinchmann}, {Charlot}, {White}, {Tremonti},
  {Kauffmann}, {Heckman}  \& {Brinkmann}}{{Brinchmann}
  et~al.}{2004}]{Brinchmann:2004lr}
{Brinchmann} J.,  {Charlot} S.,  {White} S.~D.~M.,  {Tremonti} C.,  {Kauffmann}
  G.,  {Heckman} T.,   {Brinkmann} J.,  2004, \mn@doi [\mnras]
  {10.1111/j.1365-2966.2004.07881.x}, \href
  {http://adsabs.harvard.edu/abs/2004MNRAS.351.1151B} {351, 1151}

\bibitem[\protect\citeauthoryear{{Brown}, {Martini}  \& {Andrews}}{{Brown}
  et~al.}{2016}]{Brown:2016aa}
{Brown} J.~S.,  {Martini} P.,   {Andrews} B.~H.,  2016, preprint, \href
  {http://adsabs.harvard.edu/abs/2016arXiv160201087B} {} (\mn@eprint {arXiv}
  {1602.01087})

\bibitem[\protect\citeauthoryear{{Bundy} et~al.,}{{Bundy}
  et~al.}{2015}]{Bundy:2015aa}
{Bundy} K.,  et~al., 2015, \mn@doi [\apj] {10.1088/0004-637X/798/1/7}, \href
  {http://adsabs.harvard.edu/abs/2015ApJ...798....7B} {798, 7}

\bibitem[\protect\citeauthoryear{{Calzetti}, {Kinney}  \&
  {Storchi-Bergmann}}{{Calzetti} et~al.}{1994}]{Calzetti:1994aa}
{Calzetti} D.,  {Kinney} A.~L.,   {Storchi-Bergmann} T.,  1994, \mn@doi [\apj]
  {10.1086/174346}, \href {http://adsabs.harvard.edu/abs/1994ApJ...429..582C}
  {429, 582}

\bibitem[\protect\citeauthoryear{{Campbell}, {Terlevich}  \&
  {Melnick}}{{Campbell} et~al.}{1986}]{Campbell:1986lr}
{Campbell} A.,  {Terlevich} R.,   {Melnick} J.,  1986, \mnras, \href
  {http://adsabs.harvard.edu/abs/1986MNRAS.223..811C} {223, 811}

\bibitem[\protect\citeauthoryear{{Cappellari} \& {Emsellem}}{{Cappellari} \&
  {Emsellem}}{2004}]{Cappellari:2004kx}
{Cappellari} M.,  {Emsellem} E.,  2004, \mn@doi [\pasp] {10.1086/381875}, \href
  {http://adsabs.harvard.edu/abs/2004PASP..116..138C} {116, 138}

\bibitem[\protect\citeauthoryear{{Cardelli}, {Clayton}  \& {Mathis}}{{Cardelli}
  et~al.}{1989}]{Cardelli:1989lr}
{Cardelli} J.~A.,  {Clayton} G.~C.,   {Mathis} J.~S.,  1989, \mn@doi [\apj]
  {10.1086/167900}, \href {http://adsabs.harvard.edu/abs/1989ApJ...345..245C}
  {345, 245}

\bibitem[\protect\citeauthoryear{{Cenarro}, {Cardiel}, {Gorgas}, {Peletier},
  {Vazdekis}  \& {Prada}}{{Cenarro} et~al.}{2001}]{Cenarro:2001uq}
{Cenarro} A.~J.,  {Cardiel} N.,  {Gorgas} J.,  {Peletier} R.~F.,  {Vazdekis}
  A.,   {Prada} F.,  2001, \mn@doi [\mnras] {10.1046/j.1365-8711.2001.04688.x},
  \href {http://adsabs.harvard.edu/abs/2001MNRAS.326..959C} {326, 959}

\bibitem[\protect\citeauthoryear{{Cresci}, {Mannucci}, {Sommariva}, {Maiolino},
  {Marconi}  \& {Brusa}}{{Cresci} et~al.}{2012}]{Cresci:2012aa}
{Cresci} G.,  {Mannucci} F.,  {Sommariva} V.,  {Maiolino} R.,  {Marconi} A.,
  {Brusa} M.,  2012, \mn@doi [\mnras] {10.1111/j.1365-2966.2011.20299.x}, \href
  {http://adsabs.harvard.edu/abs/2012MNRAS.421..262C} {421, 262}

\bibitem[\protect\citeauthoryear{{Croom} et~al.,}{{Croom}
  et~al.}{2012}]{Croom:2012aa}
{Croom} S.~M.,  et~al., 2012, \mn@doi [\mnras]
  {10.1111/j.1365-2966.2011.20365.x}, \href
  {http://adsabs.harvard.edu/abs/2012MNRAS.421..872C} {421, 872}

\bibitem[\protect\citeauthoryear{{Dav{\'e}}, {Finlator}  \&
  {Oppenheimer}}{{Dav{\'e}} et~al.}{2011}]{Dave:2011aa}
{Dav{\'e}} R.,  {Finlator} K.,   {Oppenheimer} B.~D.,  2011, \mn@doi [\mnras]
  {10.1111/j.1365-2966.2011.19132.x}, \href
  {http://adsabs.harvard.edu/abs/2011MNRAS.416.1354D} {416, 1354}

\bibitem[\protect\citeauthoryear{{Dayal}, {Ferrara}  \& {Dunlop}}{{Dayal}
  et~al.}{2013}]{Dayal:2013aa}
{Dayal} P.,  {Ferrara} A.,   {Dunlop} J.~S.,  2013, \mn@doi [\mnras]
  {10.1093/mnras/stt083}, \href
  {http://adsabs.harvard.edu/abs/2013MNRAS.430.2891D} {430, 2891}

\bibitem[\protect\citeauthoryear{{De Robertis}, {Dufour}  \& {Hunt}}{{De
  Robertis} et~al.}{1987}]{De-Robertis:1987rm}
{De Robertis} M.~M.,  {Dufour} R.~J.,   {Hunt} R.~W.,  1987, \jrasc, \href
  {http://adsabs.harvard.edu/abs/1987JRASC..81..195D} {81, 195}

\bibitem[\protect\citeauthoryear{{Dopita}, {Sutherland}, {Nicholls}, {Kewley}
  \& {Vogt}}{{Dopita} et~al.}{2013}]{Dopita:2013aa}
{Dopita} M.~A.,  {Sutherland} R.~S.,  {Nicholls} D.~C.,  {Kewley} L.~J.,
  {Vogt} F.~P.~A.,  2013, \mn@doi [\apjs] {10.1088/0067-0049/208/1/10}, \href
  {http://adsabs.harvard.edu/abs/2013ApJS..208...10D} {208, 10}

\bibitem[\protect\citeauthoryear{{Dopita}, {Kewley}, {Sutherland}  \&
  {Nicholls}}{{Dopita} et~al.}{2016}]{Dopita:2016aa}
{Dopita} M.~A.,  {Kewley} L.~J.,  {Sutherland} R.~S.,   {Nicholls} D.~C.,
  2016, \mn@doi [\apss] {10.1007/s10509-016-2657-8}, \href
  {http://adsabs.harvard.edu/abs/2016Ap%26SS.361...61D} {361, 61}

\bibitem[\protect\citeauthoryear{{Erb}}{{Erb}}{2006}]{Erb:2006aa}
{Erb} D.,  2006, in American Astronomical Society Meeting Abstracts. p.~925

\bibitem[\protect\citeauthoryear{{Esteban}, {Peimbert}, {Garc{\'{\i}}a-Rojas},
  {Ruiz}, {Peimbert}  \& {Rodr{\'{\i}}guez}}{{Esteban}
  et~al.}{2004}]{Esteban:2004kx}
{Esteban} C.,  {Peimbert} M.,  {Garc{\'{\i}}a-Rojas} J.,  {Ruiz} M.~T.,
  {Peimbert} A.,   {Rodr{\'{\i}}guez} M.,  2004, \mn@doi [\mnras]
  {10.1111/j.1365-2966.2004.08313.x}, \href
  {http://adsabs.harvard.edu/abs/2004MNRAS.355..229E} {355, 229}

\bibitem[\protect\citeauthoryear{{Esteban}, {Bresolin}, {Peimbert},
  {Garc{\'{\i}}a-Rojas}, {Peimbert}  \& {Mesa-Delgado}}{{Esteban}
  et~al.}{2009}]{Esteban:2009gu}
{Esteban} C.,  {Bresolin} F.,  {Peimbert} M.,  {Garc{\'{\i}}a-Rojas} J.,
  {Peimbert} A.,   {Mesa-Delgado} A.,  2009, \mn@doi [\apj]
  {10.1088/0004-637X/700/1/654}, \href
  {http://adsabs.harvard.edu/abs/2009ApJ...700..654E} {700, 654}

\bibitem[\protect\citeauthoryear{{Esteban}, {Garc{\'{\i}}a-Rojas}, {Carigi},
  {Peimbert}, {Bresolin}, {L{\'o}pez-S{\'a}nchez}  \& {Mesa-Delgado}}{{Esteban}
  et~al.}{2014}]{Esteban:2014aa}
{Esteban} C.,  {Garc{\'{\i}}a-Rojas} J.,  {Carigi} L.,  {Peimbert} M.,
  {Bresolin} F.,  {L{\'o}pez-S{\'a}nchez} A.~R.,   {Mesa-Delgado} A.,  2014,
  \mn@doi [\mnras] {10.1093/mnras/stu1177}, \href
  {http://adsabs.harvard.edu/abs/2014MNRAS.443..624E} {443, 624}

\bibitem[\protect\citeauthoryear{{Falc{\'o}n-Barroso},
  {S{\'a}nchez-Bl{\'a}zquez}, {Vazdekis}, {Ricciardelli}, {Cardiel}, {Cenarro},
  {Gorgas}  \& {Peletier}}{{Falc{\'o}n-Barroso}
  et~al.}{2011}]{Falcon-Barroso:2011qy}
{Falc{\'o}n-Barroso} J.,  {S{\'a}nchez-Bl{\'a}zquez} P.,  {Vazdekis} A.,
  {Ricciardelli} E.,  {Cardiel} N.,  {Cenarro} A.~J.,  {Gorgas} J.,
  {Peletier} R.~F.,  2011, \mn@doi [\aap] {10.1051/0004-6361/201116842}, \href
  {http://adsabs.harvard.edu/abs/2011A%26A...532A..95F} {532, A95}

\bibitem[\protect\citeauthoryear{{Garc{\'{\i}}a-Rojas} \&
  {Esteban}}{{Garc{\'{\i}}a-Rojas} \& {Esteban}}{2007}]{Garcia-Rojas:2007aa}
{Garc{\'{\i}}a-Rojas} J.,  {Esteban} C.,  2007, \mn@doi [\apj]
  {10.1086/521871}, \href {http://adsabs.harvard.edu/abs/2007ApJ...670..457G}
  {670, 457}

\bibitem[\protect\citeauthoryear{{Garnett}}{{Garnett}}{1992}]{Garnett:1992aa}
{Garnett} D.~R.,  1992, \mn@doi [\aj] {10.1086/116146}, \href
  {http://adsabs.harvard.edu/abs/1992AJ....103.1330G} {103, 1330}

\bibitem[\protect\citeauthoryear{{Izotov}, {Stasi{\'n}ska}, {Meynet}, {Guseva}
  \& {Thuan}}{{Izotov} et~al.}{2006}]{Izotov:2006aa}
{Izotov} Y.~I.,  {Stasi{\'n}ska} G.,  {Meynet} G.,  {Guseva} N.~G.,   {Thuan}
  T.~X.,  2006, \mn@doi [\aap] {10.1051/0004-6361:20053763}, \href
  {http://adsabs.harvard.edu/abs/2006A%26A...448..955I} {448, 955}

\bibitem[\protect\citeauthoryear{{Kauffmann} et~al.,}{{Kauffmann}
  et~al.}{2003a}]{Kauffmann:2003lr}
{Kauffmann} G.,  et~al., 2003a, \mn@doi [\mnras]
  {10.1046/j.1365-8711.2003.06291.x}, \href
  {http://adsabs.harvard.edu/abs/2003MNRAS.341...33K} {341, 33}

\bibitem[\protect\citeauthoryear{{Kauffmann} et~al.,}{{Kauffmann}
  et~al.}{2003b}]{Kauffmann:2003aa}
{Kauffmann} G.,  et~al., 2003b, \mn@doi [\mnras]
  {10.1111/j.1365-2966.2003.07154.x}, \href
  {http://adsabs.harvard.edu/abs/2003MNRAS.346.1055K} {346, 1055}

\bibitem[\protect\citeauthoryear{{Kennicutt}, {Bresolin}  \&
  {Garnett}}{{Kennicutt} et~al.}{2003}]{Kennicutt:2003fk}
{Kennicutt} Jr. R.~C.,  {Bresolin} F.,   {Garnett} D.~R.,  2003, \mn@doi [\apj]
  {10.1086/375398}, \href {http://adsabs.harvard.edu/abs/2003ApJ...591..801K}
  {591, 801}

\bibitem[\protect\citeauthoryear{{Kewley} \& {Dopita}}{{Kewley} \&
  {Dopita}}{2002}]{Kewley:2002aa}
{Kewley} L.~J.,  {Dopita} M.~A.,  2002, \mn@doi [\apjs] {10.1086/341326}, \href
  {http://adsabs.harvard.edu/abs/2002ApJS..142...35K} {142, 35}

\bibitem[\protect\citeauthoryear{{Kewley} \& {Ellison}}{{Kewley} \&
  {Ellison}}{2008}]{Kewley:2008aa}
{Kewley} L.~J.,  {Ellison} S.~L.,  2008, \mn@doi [\apj] {10.1086/587500}, \href
  {http://adsabs.harvard.edu/abs/2008ApJ...681.1183K} {681, 1183}

\bibitem[\protect\citeauthoryear{{Kobulnicky} \& {Kewley}}{{Kobulnicky} \&
  {Kewley}}{2004}]{Kobulnicky:2004aa}
{Kobulnicky} H.~A.,  {Kewley} L.~J.,  2004, \mn@doi [\apj] {10.1086/425299},
  \href {http://adsabs.harvard.edu/abs/2004ApJ...617..240K} {617, 240}

\bibitem[\protect\citeauthoryear{{Kobulnicky}, {Kennicutt}  \&
  {Pizagno}}{{Kobulnicky} et~al.}{1999}]{Kobulnicky:1999qy}
{Kobulnicky} H.~A.,  {Kennicutt} Jr. R.~C.,   {Pizagno} J.~L.,  1999, \mn@doi
  [\apj] {10.1086/306987}, \href
  {http://adsabs.harvard.edu/abs/1999ApJ...514..544K} {514, 544}

\bibitem[\protect\citeauthoryear{{Le Borgne}, {Rocca-Volmerange}, {Prugniel},
  {Lan{\c c}on}, {Fioc}  \& {Soubiran}}{{Le Borgne}
  et~al.}{2004}]{Le-Borgne:2004aa}
{Le Borgne} D.,  {Rocca-Volmerange} B.,  {Prugniel} P.,  {Lan{\c c}on} A.,
  {Fioc} M.,   {Soubiran} C.,  2004, \mn@doi [\aap]
  {10.1051/0004-6361:200400044}, \href
  {http://adsabs.harvard.edu/abs/2004A%26A...425..881L} {425, 881}

\bibitem[\protect\citeauthoryear{{Lee}, {Skillman}, {Cannon}, {Jackson},
  {Gehrz}, {Polomski}  \& {Woodward}}{{Lee} et~al.}{2006}]{Lee:2006aa}
{Lee} H.,  {Skillman} E.~D.,  {Cannon} J.~M.,  {Jackson} D.~C.,  {Gehrz} R.~D.,
   {Polomski} E.~F.,   {Woodward} C.~E.,  2006, \mn@doi [\apj]
  {10.1086/505573}, \href {http://adsabs.harvard.edu/abs/2006ApJ...647..970L}
  {647, 970}

\bibitem[\protect\citeauthoryear{{Liang}, {Hammer}, {Yin}, {Flores},
  {Rodrigues}  \& {Yang}}{{Liang} et~al.}{2007}]{Liang:2007aa}
{Liang} Y.~C.,  {Hammer} F.,  {Yin} S.~Y.,  {Flores} H.,  {Rodrigues} M.,
  {Yang} Y.~B.,  2007, \mn@doi [\aap] {10.1051/0004-6361:20077436}, \href
  {http://adsabs.harvard.edu/abs/2007A%26A...473..411L} {473, 411}

\bibitem[\protect\citeauthoryear{{Lilly}, {Carollo}, {Pipino}, {Renzini}  \&
  {Peng}}{{Lilly} et~al.}{2013}]{Lilly:2013aa}
{Lilly} S.~J.,  {Carollo} C.~M.,  {Pipino} A.,  {Renzini} A.,   {Peng} Y.,
  2013, \mn@doi [\apj] {10.1088/0004-637X/772/2/119}, \href
  {http://adsabs.harvard.edu/abs/2013ApJ...772..119L} {772, 119}

\bibitem[\protect\citeauthoryear{{L{\'o}pez-S{\'a}nchez}, {Dopita}, {Kewley},
  {Zahid}, {Nicholls}  \& {Scharw{\"a}chter}}{{L{\'o}pez-S{\'a}nchez}
  et~al.}{2012}]{Lopez-Sanchez:2012aa}
{L{\'o}pez-S{\'a}nchez} {\'A}.~R.,  {Dopita} M.~A.,  {Kewley} L.~J.,  {Zahid}
  H.~J.,  {Nicholls} D.~C.,   {Scharw{\"a}chter} J.,  2012, \mn@doi [\mnras]
  {10.1111/j.1365-2966.2012.21145.x}, \href
  {http://adsabs.harvard.edu/abs/2012MNRAS.426.2630L} {426, 2630}

\bibitem[\protect\citeauthoryear{{Luridiana}, {Morisset}  \&
  {Shaw}}{{Luridiana} et~al.}{2012}]{Luridiana:2012aa}
{Luridiana} V.,  {Morisset} C.,   {Shaw} R.~A.,  2012, in IAU Symposium. pp
  422--423, \mn@doi{10.1017/S1743921312011738}

\bibitem[\protect\citeauthoryear{{Luridiana}, {Morisset}  \&
  {Shaw}}{{Luridiana} et~al.}{2015}]{Luridiana:2015aa}
{Luridiana} V.,  {Morisset} C.,   {Shaw} R.~A.,  2015, \mn@doi [\aap]
  {10.1051/0004-6361/201323152}, \href
  {http://adsabs.harvard.edu/abs/2015A%26A...573A..42L} {573, A42}

\bibitem[\protect\citeauthoryear{{Maiolino} et~al.,}{{Maiolino}
  et~al.}{2008}]{Maiolino:2008cv}
{Maiolino} R.,  et~al., 2008, \mn@doi [\aap] {10.1051/0004-6361:200809678},
  \href {http://adsabs.harvard.edu/abs/2008A%26A...488..463M} {488, 463}

\bibitem[\protect\citeauthoryear{{Mannucci} et~al.,}{{Mannucci}
  et~al.}{2009}]{Mannucci:2009aa}
{Mannucci} F.,  et~al., 2009, \mn@doi [\mnras]
  {10.1111/j.1365-2966.2009.15185.x}, \href
  {http://adsabs.harvard.edu/abs/2009MNRAS.398.1915M} {398, 1915}

\bibitem[\protect\citeauthoryear{{Mannucci}, {Cresci}, {Maiolino}, {Marconi}
  \& {Gnerucci}}{{Mannucci} et~al.}{2010}]{Mannucci:2010gy}
{Mannucci} F.,  {Cresci} G.,  {Maiolino} R.,  {Marconi} A.,   {Gnerucci} A.,
  2010, \mn@doi [\mnras] {10.1111/j.1365-2966.2010.17291.x}, \href
  {http://adsabs.harvard.edu/abs/2010MNRAS.408.2115M} {408, 2115}

\bibitem[\protect\citeauthoryear{{Marino} et~al.,}{{Marino}
  et~al.}{2013}]{Marino:2013ty}
{Marino} R.~A.,  et~al., 2013, \mn@doi [\aap] {10.1051/0004-6361/201321956},
  \href {http://adsabs.harvard.edu/abs/2013A%26A...559A.114M} {559, A114}

\bibitem[\protect\citeauthoryear{{Masters}, {Faisst}  \& {Capak}}{{Masters}
  et~al.}{2016}]{Masters:2016aa}
{Masters} D.,  {Faisst} A.,   {Capak} P.,  2016, preprint, \href
  {http://adsabs.harvard.edu/abs/2016arXiv160504314M} {} (\mn@eprint {arXiv}
  {1605.04314})

\bibitem[\protect\citeauthoryear{{McGaugh}}{{McGaugh}}{1991}]{McGaugh:1991aa}
{McGaugh} S.~S.,  1991, \mn@doi [\apj] {10.1086/170569}, \href
  {http://adsabs.harvard.edu/abs/1991ApJ...380..140M} {380, 140}

\bibitem[\protect\citeauthoryear{{Moustakas} \& {Kennicutt}}{{Moustakas} \&
  {Kennicutt}}{2006}]{Moustakas:2006aa}
{Moustakas} J.,  {Kennicutt} Jr. R.~C.,  2006, \mn@doi [\apjs]
  {10.1086/500971}, \href {http://adsabs.harvard.edu/abs/2006ApJS..164...81M}
  {164, 81}

\bibitem[\protect\citeauthoryear{{Moustakas}, {Kennicutt}, {Tremonti}, {Dale},
  {Smith}  \& {Calzetti}}{{Moustakas} et~al.}{2010}]{Moustakas:2010aa}
{Moustakas} J.,  {Kennicutt} Jr. R.~C.,  {Tremonti} C.~A.,  {Dale} D.~A.,
  {Smith} J.-D.~T.,   {Calzetti} D.,  2010, \mn@doi [\apjs]
  {10.1088/0067-0049/190/2/233}, \href
  {http://adsabs.harvard.edu/abs/2010ApJS..190..233M} {190, 233}

\bibitem[\protect\citeauthoryear{{Nagao}, {Maiolino}  \& {Marconi}}{{Nagao}
  et~al.}{2006}]{Nagao:2006gd}
{Nagao} T.,  {Maiolino} R.,   {Marconi} A.,  2006, \mn@doi [\aap]
  {10.1051/0004-6361:20065216}, \href
  {http://adsabs.harvard.edu/abs/2006A%26A...459...85N} {459, 85}

\bibitem[\protect\citeauthoryear{{Nicholls}, {Dopita}  \&
  {Sutherland}}{{Nicholls} et~al.}{2012}]{Nicholls:2012aa}
{Nicholls} D.~C.,  {Dopita} M.~A.,   {Sutherland} R.~S.,  2012, \mn@doi [\apj]
  {10.1088/0004-637X/752/2/148}, \href
  {http://adsabs.harvard.edu/abs/2012ApJ...752..148N} {752, 148}

\bibitem[\protect\citeauthoryear{{Nicholls}, {Dopita}, {Sutherland}, {Kewley}
  \& {Palay}}{{Nicholls} et~al.}{2013}]{Nicholls:2013aa}
{Nicholls} D.~C.,  {Dopita} M.~A.,  {Sutherland} R.~S.,  {Kewley} L.~J.,
  {Palay} E.,  2013, \mn@doi [\apjs] {10.1088/0067-0049/207/2/21}, \href
  {http://adsabs.harvard.edu/abs/2013ApJS..207...21N} {207, 21}

\bibitem[\protect\citeauthoryear{{Noeske} et~al.,}{{Noeske}
  et~al.}{2007}]{Noeske:2007aa}
{Noeske} K.~G.,  et~al., 2007, \mn@doi [\apjl] {10.1086/517927}, \href
  {http://adsabs.harvard.edu/abs/2007ApJ...660L..47N} {660, L47}

\bibitem[\protect\citeauthoryear{{Pagel}, {Edmunds}, {Blackwell}, {Chun}  \&
  {Smith}}{{Pagel} et~al.}{1979}]{Pagel:1979pd}
{Pagel} B.~E.~J.,  {Edmunds} M.~G.,  {Blackwell} D.~E.,  {Chun} M.~S.,
  {Smith} G.,  1979, \mnras, \href
  {http://adsabs.harvard.edu/abs/1979MNRAS.189...95P} {189, 95}

\bibitem[\protect\citeauthoryear{{Palay}, {Nahar}, {Pradhan}  \&
  {Eissner}}{{Palay} et~al.}{2012}]{Palay:2012aa}
{Palay} E.,  {Nahar} S.~N.,  {Pradhan} A.~K.,   {Eissner} W.,  2012, \mn@doi
  [\mnras] {10.1111/j.1745-3933.2012.01252.x}, \href
  {http://adsabs.harvard.edu/abs/2012MNRAS.423L..35P} {423, L35}

\bibitem[\protect\citeauthoryear{{Peimbert}}{{Peimbert}}{1967}]{Peimbert:1967qv}
{Peimbert} M.,  1967, \mn@doi [\apj] {10.1086/149385}, \href
  {http://adsabs.harvard.edu/abs/1967ApJ...150..825P} {150, 825}

\bibitem[\protect\citeauthoryear{{Peimbert} \& {Peimbert}}{{Peimbert} \&
  {Peimbert}}{2014}]{Peimbert:2014aa}
{Peimbert} M.,  {Peimbert} A.,  2014, in Revista Mexicana de Astronomia y
  Astrofisica Conference Series. pp 137--137

\bibitem[\protect\citeauthoryear{{Peimbert}, {Peimbert}, {Esteban},
  {Garc{\'{\i}}a-Rojas}, {Bresolin}, {Carigi}, {Ruiz}  \&
  {L{\'o}pez-S{\'a}nchez}}{{Peimbert} et~al.}{2007}]{Peimbert:2007aa}
{Peimbert} M.,  {Peimbert} A.,  {Esteban} C.,  {Garc{\'{\i}}a-Rojas} J.,
  {Bresolin} F.,  {Carigi} L.,  {Ruiz} M.~T.,   {L{\'o}pez-S{\'a}nchez} A.~R.,
  2007, in {Guzm{\'a}n} R.,  ed.,  Revista Mexicana de Astronomia y Astrofisica
  Conference Series Vol. 29, Revista Mexicana de Astronomia y Astrofisica
  Conference Series. pp 72--79 (\mn@eprint {} {astro-ph/0608440})

\bibitem[\protect\citeauthoryear{{Pettini} \& {Pagel}}{{Pettini} \&
  {Pagel}}{2004}]{Pettini:2004fk}
{Pettini} M.,  {Pagel} B.~E.~J.,  2004, \mn@doi [\mnras]
  {10.1111/j.1365-2966.2004.07591.x}, \href
  {http://adsabs.harvard.edu/abs/2004MNRAS.348L..59P} {348, L59}

\bibitem[\protect\citeauthoryear{{Pilyugin}}{{Pilyugin}}{2005}]{Pilyugin:2005aa}
{Pilyugin} L.~S.,  2005, \mn@doi [\aap] {10.1051/0004-6361:200500108}, \href
  {http://adsabs.harvard.edu/abs/2005A%26A...436L...1P} {436, L1}

\bibitem[\protect\citeauthoryear{{Pilyugin}}{{Pilyugin}}{2006}]{Pilyugin:2007}
{Pilyugin} L.~S.,  2006, \mnras, 375, 685

\bibitem[\protect\citeauthoryear{{Pilyugin} \& {Grebel}}{{Pilyugin} \&
  {Grebel}}{2016}]{Pilyugin:2016aa}
{Pilyugin} L.~S.,  {Grebel} E.~K.,  2016, \mn@doi [\mnras]
  {10.1093/mnras/stw238}, \href
  {http://adsabs.harvard.edu/abs/2016MNRAS.tmp...29P} {}

\bibitem[\protect\citeauthoryear{{Pilyugin} \& {Thuan}}{{Pilyugin} \&
  {Thuan}}{2005}]{Pilyugin:2005}
{Pilyugin} L.~S.,  {Thuan} T.~X.,  2005, Astrophysical Journal, 631, 231

\bibitem[\protect\citeauthoryear{{Pilyugin}, {Thuan}  \&
  {V{\'{\i}}lchez}}{{Pilyugin} et~al.}{2006a}]{Pilyugin:2006ab}
{Pilyugin} L.~S.,  {Thuan} T.~X.,   {V{\'{\i}}lchez} J.~M.,  2006a, \mn@doi
  [\mnras] {10.1111/j.1365-2966.2006.10033.x}, \href
  {http://adsabs.harvard.edu/abs/2006MNRAS.367.1139P} {367, 1139}

\bibitem[\protect\citeauthoryear{{Pilyugin}, {V{\'{\i}}lchez}  \&
  {Thuan}}{{Pilyugin} et~al.}{2006b}]{Pilyugin:2006aa}
{Pilyugin} L.~S.,  {V{\'{\i}}lchez} J.~M.,   {Thuan} T.~X.,  2006b, \mn@doi
  [\mnras] {10.1111/j.1365-2966.2006.10618.x}, \href
  {http://adsabs.harvard.edu/abs/2006MNRAS.370.1928P} {370, 1928}

\bibitem[\protect\citeauthoryear{{Pilyugin}, {Mattsson}, {V{\'{\i}}lchez}  \&
  {Cedr{\'e}s}}{{Pilyugin} et~al.}{2009}]{Pilyugin:2009aa}
{Pilyugin} L.~S.,  {Mattsson} L.,  {V{\'{\i}}lchez} J.~M.,   {Cedr{\'e}s} B.,
  2009, \mn@doi [\mnras] {10.1111/j.1365-2966.2009.15182.x}, \href
  {http://adsabs.harvard.edu/abs/2009MNRAS.398..485P} {398, 485}

\bibitem[\protect\citeauthoryear{{Pilyugin}, {V{\'{\i}}lchez}, {Cedr{\'e}s}  \&
  {Thuan}}{{Pilyugin} et~al.}{2010a}]{Pilyugin:2010ab}
{Pilyugin} L.~S.,  {V{\'{\i}}lchez} J.~M.,  {Cedr{\'e}s} B.,   {Thuan} T.~X.,
  2010a, \mn@doi [\mnras] {10.1111/j.1365-2966.2009.16166.x}, \href
  {http://adsabs.harvard.edu/abs/2010MNRAS.403..896P} {403, 896}

\bibitem[\protect\citeauthoryear{{Pilyugin}, {V{\'{\i}}lchez}  \&
  {Thuan}}{{Pilyugin} et~al.}{2010b}]{Pilyugin:2010aa}
{Pilyugin} L.~S.,  {V{\'{\i}}lchez} J.~M.,   {Thuan} T.~X.,  2010b, \mn@doi
  [\apj] {10.1088/0004-637X/720/2/1738}, \href
  {http://adsabs.harvard.edu/abs/2010ApJ...720.1738P} {720, 1738}

\bibitem[\protect\citeauthoryear{{Pilyugin}, {Grebel}  \&
  {Mattsson}}{{Pilyugin} et~al.}{2012}]{Pilyugin:2012_2}
{Pilyugin} L.~S.,  {Grebel} E.~K.,   {Mattsson} L.,  2012, \mnras, 424, 2316

\bibitem[\protect\citeauthoryear{{Ricciardelli}, {Vazdekis}, {Cenarro}  \&
  {Falc{\'o}n-Barroso}}{{Ricciardelli} et~al.}{2012}]{Ricciardelli:2012aa}
{Ricciardelli} E.,  {Vazdekis} A.,  {Cenarro} A.~J.,   {Falc{\'o}n-Barroso} J.,
   2012, \mn@doi [\mnras] {10.1111/j.1365-2966.2012.21178.x}, \href
  {http://adsabs.harvard.edu/abs/2012MNRAS.424..172R} {424, 172}

\bibitem[\protect\citeauthoryear{{Salim} et~al.,}{{Salim}
  et~al.}{2007}]{Salim:2007aa}
{Salim} S.,  et~al., 2007, \mn@doi [\apjs] {10.1086/519218}, \href
  {http://adsabs.harvard.edu/abs/2007ApJS..173..267S} {173, 267}

\bibitem[\protect\citeauthoryear{{S{\'a}nchez-Bl{\'a}zquez}
  et~al.,}{{S{\'a}nchez-Bl{\'a}zquez} et~al.}{2006}]{Sanchez-Blazquez:2006fk}
{S{\'a}nchez-Bl{\'a}zquez} P.,  et~al., 2006, \mn@doi [\mnras]
  {10.1111/j.1365-2966.2006.10699.x}, \href
  {http://adsabs.harvard.edu/abs/2006MNRAS.371..703S} {371, 703}

\bibitem[\protect\citeauthoryear{{S{\'a}nchez} et~al.,}{{S{\'a}nchez}
  et~al.}{2012}]{Sanchez:2012aa}
{S{\'a}nchez} S.~F.,  et~al., 2012, \mn@doi [\aap]
  {10.1051/0004-6361/201117353}, \href
  {http://adsabs.harvard.edu/abs/2012A%26A...538A...8S} {538, A8}

\bibitem[\protect\citeauthoryear{{Stasi{\'n}ska}}{{Stasi{\'n}ska}}{2002}]{Stasinska:2002lr}
{Stasi{\'n}ska} G.,  2002, ArXiv Astrophysics e-prints, \href
  {http://adsabs.harvard.edu/abs/2002astro.ph..7500S} {}

\bibitem[\protect\citeauthoryear{{Stasi{\'n}ska}}{{Stasi{\'n}ska}}{2005}]{Stasinska:2005aa}
{Stasi{\'n}ska} G.,  2005, \mn@doi [\aap] {10.1051/0004-6361:20042216}, \href
  {http://adsabs.harvard.edu/abs/2005A%26A...434..507S} {434, 507}

\bibitem[\protect\citeauthoryear{{Tremonti} et~al.,}{{Tremonti}
  et~al.}{2004}]{Tremonti:2004aa}
{Tremonti} C.~A.,  et~al., 2004, \mn@doi [\apj] {10.1086/423264}, \href
  {http://adsabs.harvard.edu/abs/2004ApJ...613..898T} {613, 898}

\bibitem[\protect\citeauthoryear{{Troncoso} et~al.,}{{Troncoso}
  et~al.}{2014}]{Troncoso:2014aa}
{Troncoso} P.,  et~al., 2014, \mn@doi [\aap] {10.1051/0004-6361/201322099},
  \href {http://adsabs.harvard.edu/abs/2014A%26A...563A..58T} {563, A58}

\bibitem[\protect\citeauthoryear{{Vazdekis}, {S{\'a}nchez-Bl{\'a}zquez},
  {Falc{\'o}n-Barroso}, {Cenarro}, {Beasley}, {Cardiel}, {Gorgas}  \&
  {Peletier}}{{Vazdekis} et~al.}{2010}]{Vazdekis:2010lr}
{Vazdekis} A.,  {S{\'a}nchez-Bl{\'a}zquez} P.,  {Falc{\'o}n-Barroso} J.,
  {Cenarro} A.~J.,  {Beasley} M.~A.,  {Cardiel} N.,  {Gorgas} J.,   {Peletier}
  R.~F.,  2010, \mn@doi [\mnras] {10.1111/j.1365-2966.2010.16407.x}, \href
  {http://adsabs.harvard.edu/abs/2010MNRAS.404.1639V} {404, 1639}

\bibitem[\protect\citeauthoryear{{Vazdekis}, {Ricciardelli}, {Cenarro},
  {Rivero-Gonz{\'a}lez}, {D{\'{\i}}az-Garc{\'{\i}}a}  \&
  {Falc{\'o}n-Barroso}}{{Vazdekis} et~al.}{2012}]{Vazdekis:2012aa}
{Vazdekis} A.,  {Ricciardelli} E.,  {Cenarro} A.~J.,  {Rivero-Gonz{\'a}lez}
  J.~G.,  {D{\'{\i}}az-Garc{\'{\i}}a} L.~A.,   {Falc{\'o}n-Barroso} J.,  2012,
  \mn@doi [\mnras] {10.1111/j.1365-2966.2012.21179.x}, \href
  {http://adsabs.harvard.edu/abs/2012MNRAS.424..157V} {424, 157}

\bibitem[\protect\citeauthoryear{{Zahid}, {Bresolin}, {Kewley}, {Coil}  \&
  {Dav{\'e}}}{{Zahid} et~al.}{2012}]{Zahid:2012aa}
{Zahid} H.~J.,  {Bresolin} F.,  {Kewley} L.~J.,  {Coil} A.~L.,   {Dav{\'e}} R.,
   2012, \mn@doi [\apj] {10.1088/0004-637X/750/2/120}, \href
  {http://adsabs.harvard.edu/abs/2012ApJ...750..120Z} {750, 120}

\bibitem[\protect\citeauthoryear{{Zaritsky}, {Kennicutt}  \&
  {Huchra}}{{Zaritsky} et~al.}{1994}]{Zaritsky:1994aa}
{Zaritsky} D.,  {Kennicutt} Jr. R.~C.,   {Huchra} J.~P.,  1994, \mn@doi [\apj]
  {10.1086/173544}, \href {http://adsabs.harvard.edu/abs/1994ApJ...420...87Z}
  {420, 87}

\makeatother
\end{thebibliography}

\input{metal_calibrations_clean.bbl}

\end{document}